\documentclass[12pt]{article}

\usepackage[normalem]{ulem}
\usepackage[mathscr]{eucal}
\usepackage[version=3]{mhchem}

\usepackage{epsfig,latexsym,amsfonts,amsmath,amsthm,amssymb,amsbsy,multirow,slashed,wasysym,textcomp,dsfont,comment,mathtools,cancel,cite,datetime,appendix,BOONDOX-calo}
\usepackage{tocloft}
\usepackage{bbold}
\usepackage{tikz}
\usetikzlibrary{backgrounds}
\usetikzlibrary{patterns}
\usetikzlibrary{arrows.meta}
\tikzstyle{bag} = [align=center]
\usetikzlibrary{decorations.pathmorphing}
\usetikzlibrary{bending}

\def\bea{\begin{eqnarray}}
\def\eea{\end{eqnarray}}

\usepackage{hyperref}
\usepackage{subcaption}

 \newcommand{\badat}{\begin{alignedat}}
 \newcommand{\eadat}{\end{alignedat}}
 
 \def\be{\begin{equation}}
\def\ee{\end{equation}}

\usepackage{xcolor}

\newcommand{\pink}[1]{\textcolor{\pink}{#1}}

\definecolor{dblue}{rgb}{0.2,0.50,0.80}

\def\bh{{\bar h}}
\def\bz{{\bar z}}
\def\bw{{\bar w}}

\def\n{k}

\def\bh{{\bar h}}
\def\bz{{\bar z}}
\def\bw{{\bar w}}

\def\pa{{\partial}}
\def\a{{\alpha}}
\def\D{{\Delta}}
\def\d{{\delta}}
\def\o{{\omega}}
\def\b{{\beta}}
\def\g{{\gamma}}

\def\n{{\nu}}

\def\e{{\epsilon}}

\usepackage[retainorgcmds]{IEEEtrantools}
\def\bea{\begin{IEEEeqnarray*}}
\def\eea{\end{IEEEeqnarray*}}
\def\n{\IEEEyesnumber}
\def\sn{\IEEEyessubnumber}

\DeclareMathOperator*{\Res}{Res}

\DeclareFontFamily{OT1}{pzc}{}
\DeclareFontShape{OT1}{pzc}{m}{it}{<-> s * [1.10] pzcmi7t}{}
\DeclareMathAlphabet{\mathpzc}{OT1}{pzc}{m}{it}

\definecolor{vert}{rgb}{0.1367 0.543 0.1367}

\numberwithin{equation}{section} 

\begin{document}

 \begin{titlepage}
  \thispagestyle{empty}
  \begin{flushright}
  \end{flushright}
  \bigskip
  \begin{center}

        \baselineskip=13pt {\LARGE \scshape{
       Detector Operators \\[.5em] for Celestial Symmetries
     }}
      \vskip1cm 

   \centerline{ 
   {Yangrui Hu} 
    and {Sabrina Pasterski}
}

\bigskip\bigskip
 
 \centerline{
\it Perimeter Institute for Theoretical Physics, Waterloo, ON N2L 2Y5, Canada}

\bigskip\bigskip

\end{center}

\begin{abstract}
 \noindent 
 This paper presents a systematic cataloging of the generators of celestial symmetries on phase space. Starting from the celestial OPEs, we first show how to extract a representation of the general-spin analog of the wedge subalgebra of $w_{1+\infty}$ on the phase space of massless matter fields of arbitrary helicity. These generators can be expressed as light-sheet operators that are quadratic in the matter fields at future or past null infinity. We next show how to extend these symmetries beyond the wedge. Doing so requires us to augment the quadratic operators with: 1) linear terms corresponding to primary descendants of the negative helicity gauge fields the matter modes couple to, and 2) a tower of higher-particle composite operator contributions. These modes can be realized as light-ray operators supported on generators of null infinity, but local on the celestial sphere. Finally, we construct a representation of the celestial symmetries that captures how the positive helicity gauge fields transform. We close by discussing how these celestial symmetries inform our choice of detector operators.

\end{abstract}

\end{titlepage}

\tableofcontents

\section{Introduction}\label{sec:intro}

The celestial holography program proposes a duality between quantum gravity in asymptotically flat spacetime and a CFT living on the codimension-two celestial sphere~\cite{Pasterski:2021rjz,Pasterski:2021raf,Raclariu:2021zjz}. A compelling feature of this celestial CFT (CCFT) is that it contains a large number of currents~\cite{Guevara:2021abz,Strominger:2021mtt} that in the bulk dual arise from the asymptotic symmetry group~\cite{Bondi:1962px,Sachs:1962wk, Sachs:1962zza,Strominger:2017zoo,Fotopoulos:2019vac,Fotopoulos:2020bqj} and the structure of the collinear limits of scattering~\cite{Fan:2019emx,Pate:2019lpp,Himwich:2021dau}. If we focus on a bottom-up approach to a flat hologram -- centered around matching symmetries of the bulk and boundary -- we can ask how these currents can help us organize the scattering matrix. In particular: what set of detector operators should we use to make measurements at null infinity?

In the previous work~\cite{Hu:2022txx} we incorporated matter fields into the asymptotic phase space representation~\cite{Freidel:2021ytz} of the $w_{1+\infty}$ symmetries found for perturbative gravity in~\cite{Strominger:2021mtt,Guevara:2021abz}, and showed that these give a generalization of the detector operators~\cite{Caron-Huot:2022eqs} that appear in the conformal collider literature~\cite{Hofman:2008ar,Belitsky:2013ofa,Belitsky:2013bja,Belitsky:2013xxa,Kravchuk:2018htv,Cordova:2018ygx,Dixon:2019uzg,Kologlu:2019mfz,Chen:2019bpb,Chang:2020qpj,Gonzo:2020xza,Chang:2022ryc,Lee:2022ige,Chen:2022jhb,Chang:2022ryc}. The aim was to use this as a step towards merging celestial holography and conformal collider programs, both of which rely on universal features of the correlation functions of boost-primary operators~\cite{Pasterski:2016qvg,Pasterski:2017kqt,Pasterski:2017ylz} at the conformal boundary to constrain scattering.

The goal of this paper is to systematically catalog the detector operators that appear as celestial symmetry generators on the radiative phase space in 4D. In doing so, we find that we can: construct purely matter representations of the wedge subalgebras, generalize the results presented for spin-2 and the $Lw_{1+\infty}$ symmetry in~\cite{Freidel:2021ytz,Hu:2022txx} to arbitrary spin, and incorporate the opposite helicity modes. The first point ties closely into the conformal collider literature because we are asking if we can realize the celestial symmetries purely in the matter sector, before coupling to any gauge fields. The fact that these detector operators are supported on the full light sheet as opposed to a single light ray explains why these enhanced symmetries would not have been seen in the cataloging of Cordova-Shao and others~\cite{Cordova:2018ygx,Belin:2020lsr}. Indeed, the algebra of quadratic light-ray operators appearing in the celestial $w_{1+\infty}$ would not close outside the wedge unless we turn on the coupling to the gravity and introduce linear and higher-order operators~\cite{Hu:2022txx}. While we will return to more detailed investigations in the spirit of the conformal collider and detector operator literature in future work, here our aim is to understand how the work by Freidel et al. \cite{Freidel:2021ytz,Freidel:2023gue} maps into the conformal collider story. A nice by-product is that our methodology sheds light on the role of the equations of motion in Freidel et al.'s story. We don't need them to talk about the operator algebras, but only when trying to phrase it as a symmetry of the $\cal S$-matrix connecting $in$ and $out$.

Our procedure is as follows: starting from the anti-holomorphic collinear singularity one can extract a chiral symmetry algebra for a tower of soft currents, defined as the residues at (negative) integer conformal dimensions of any massless gauge field. These fall into finite-dimensional multiplets of the left-handed $SL(2)$ of the complexified Lorentz group. The celestial symmetry algebra follows from treating these modes as functions of the anti-holomorphic celestial sphere coordinate $\bz$ and computing complexified radial-ordered commutators. The point of \cite{Freidel:2021ytz} was to look for a homomorphism of this algebra that maps the radially ordered bracket to the symplectic product on phase space 
\be\label{bracket}
\big[A(z,\bz),B(w,\bw)\big]_{\bw}=\oint_{\bw}\frac{d\bz}{2\pi i}\,A(z,\bz)\,B(w,\bw) ~~\mapsto ~~  i\hbar\,[A,B]=\{A,B\}_{P.B.} ~~. 
\ee
So the goal now is to construct operators that obey the celestial symmetries via the standard phase space bracket, as well as operators that transform in the same representations of these symmetries as the gauge fields from the opposite helicity sector.

As observed in~\cite{Himwich:2021dau}, the collinear conformally soft theorems imply that the action of the celestial symmetries on matter fields acts like a vector field on the celestial sphere directions while simultaneously shifting the weights.\footnote{The non-locality in the $u$ can be tamed in the integer basis~\cite{Freidel:2022skz,Freidel:2023gue}.} We can use this simple charge action to straightforwardly construct quadratic operators that realize the celestial symmetries on massless matter fields. As compared to~\cite{Freidel:2023gue} we get the matter sector ``hard" charges for the celestial symmetries. These only close within the wedge, which can be seen by examining the Jacobi identity.

\begin{figure}[t]
\centering
\begin{tikzpicture}[scale=1]
\definecolor{darkgreen}{rgb}{.0, 0.5, .1};
\draw[white, fill= cyan!20!white ] (1,0) -- (-2,3)--(-2,-3) -- (1,0);
\fill[red!20!white ] (0,-3) -- (2/2-.05,-3)--(2/2-.05,3) -- (0,3) -- (0,-3);
\fill[purple!40!white ](0,-1) -- (2/2-.05,0) -- (0,1) -- (0,-1);
\draw[->,thick] (-2,0) -- (2,0) node[right]{$\Delta$};
\draw[->,thick] (0,-3) -- (0,3) node[above] {$n$}; 
\draw[] (1,0) -- (-2,3);
\draw[] (1,0) -- (-2,-3);
\draw[black,fill=white] (1,0) circle (2pt);
\filldraw[black] (1/2,1/2) circle (2pt);
\filldraw[black] (1/2,-1/2) circle (2pt);
\filldraw[black] (0,1) circle (2pt);
\filldraw[black] (0,0) circle (2pt);
\filldraw[black] (0,-1) circle (2pt);
\filldraw[black] (-1/2,1/2) circle (1pt);
\filldraw[black] (-1/2,-1/2) circle (1pt);
\filldraw[black] (-1/2,3/2) circle (1pt);
\filldraw[black] (-1/2,-3/2) circle (1pt);
\filldraw[black] (-1,2) circle (1pt);
\filldraw[black] (-1,1) circle (1pt);
\filldraw[black] (-1,0) circle (1pt);
\filldraw[black] (-1,-1) circle (1pt);
\filldraw[black] (-1,-2) circle (1pt);
\filldraw[black] (-3/2,-5/2) circle (1pt);
\filldraw[black] (-3/2,-3/2) circle (1pt);
\filldraw[black] (-3/2,-1/2) circle (1pt);
\filldraw[black] (-3/2,1/2) circle (1pt);
\filldraw[black] (-3/2,3/2) circle (1pt);
\filldraw[black] (-3/2,5/2) circle (1pt);
\filldraw[black] (1/2,3/2) circle (0.7pt);
\filldraw[black] (1/2,-3/2) circle (0.7pt);
\filldraw[black] (1/2,5/2) circle (0.7pt);
\filldraw[black] (1/2,-5/2) circle (0.7pt);
\filldraw[black] (0,2) circle (0.7pt);
\filldraw[black] (0,-2) circle (0.7pt);
\filldraw[black] (-1/2,5/2) circle (0.7pt);
\filldraw[black] (-1/2,-5/2) circle (0.7pt);
\filldraw[black] (-1,3) circle (0.7pt);
\filldraw[black] (-1,-3) circle (0.7pt);
\draw[black,fill=white] (1,1) circle (0.7pt);
\draw[black,fill=white] (1,-1) circle (0.7pt);
\draw[black,fill=white] (1,2) circle (0.7pt);
\draw[black,fill=white] (1,-2) circle (0.7pt);
\draw[black,fill=white] (1,3) circle (0.7pt);
\draw[black,fill=white] (1,-3) circle (0.7pt);
\draw[black,fill=white] (3/2,1/2) circle (0.7pt);
\draw[black,fill=white] (3/2,-1/2) circle (0.7pt);
\draw[black,fill=white] (3/2,3/2) circle (0.7pt);
\draw[black,fill=white] (3/2,-3/2) circle (0.7pt);
\draw[black,fill=white] (3/2,5/2) circle (0.7pt);
\draw[black,fill=white] (3/2,-5/2) circle (0.7pt);
\end{tikzpicture}

\caption{Here we plot the modes of the negative helicity conformally soft gravitons corresponding to the $w_{1+\infty}$ symmetry in celestial CFT. The horizontal axis $\Delta$ corresponds to the boost weight of the soft graviton mode, picked out as a residue in the complex $\Delta$ plane. The soft limits of amplitudes imply that residues to the right of the principal series vanish (open dots). The celestial symmetry algebras are extracted from the anti-holomorphic collinear singularities we get when we complexify the celestial sphere. The vertical axis corresponds to projecting onto powers of $z$ via the integral $\oint dz z^{n+\frac{\D}{2}-2}$.  Symmetry `charges' are picked out via an additional $\oint d\bz$ integral~\cite{Stieberger:2018onx,Banerjee:2020kaa,Himwich:2021dau}.
The wedge subalgebra is bounded by $n=\pm \frac{2-\D}{2}$ (indicated in blue), while the BMS operators are in the vertical strip corresponding to the $\Delta=1$ and $\Delta=0$ modes (shaded in pink). The intersection contains the five $\chi$-Poincaré generators, which are denoted by larger dots surrounding the purple region.
\label{fig:diamonds-mode}}
\end{figure}
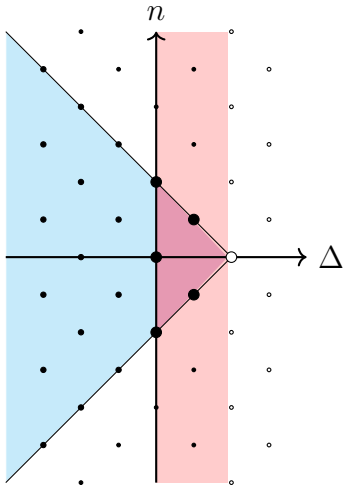

Projecting onto the wedge involves smearing our light-ray operators on the full light sheet. Extending the symmetries outside the wedge amounts to being able to realize them in terms of light-ray operators that are local on the celestial sphere. However, to do so, we need to couple the matter to the gauge fields and add multi-particle terms. Schematically, the total charge $Q$ can be decomposed as follows.
\begin{equation}\label{Qk}
    Q ~=~ Q^1 ~+~ Q^2 ~+~ Q^3 ~+~ \cdots ~~,
\end{equation}
where the superscript labels the number of fields. 
We can take the leading term to be the ${\rm SL}(2)_L$ primary descendant of a corresponding $u$-moment of the spin-$s$ gauge fields~\cite{Pasterski:2021dqe}, which forms a tower of higher-spin dimension $\Delta=1+s$ operators.

One can check that the truncation of the operator algebra at linear order agrees with the celestial chiral symmetry algebra while matching up to quadratic order determines the cubic terms and so forth. Matching the collinear couplings to other matter fields requires adding additional quadratic terms in the same spirit as \cite{Hu:2022txx}. Meanwhile, the same procedure can be applied to the mixed helicity celestial OPE to get a set of higher-spin dimension $\Delta=1-s$ operators.
In this work, we will stick to massless matter fields and integer spins. For the symmetries associated with half-integer spins, for example supersymmetry, one can construct their representations in phase space in a similar manner. While we leave the generalization to incorporate massive matter fields to future work.

To drive home the different notions of infinite dimensional symmetry enhancements we will encounter herein, we've illustrated three distinct closed subalgebras for gravity in figure~\ref{fig:diamonds-mode}.
The infinite-dimensional symmetry enhancements in~\cite{Strominger:2017zoo} are the angle-dependent enhancements of the translations and rotations to the BMS group. These correspond to a tower of $z$-modes at $\D=1$ and 0 and are in the region shaded in pink in figure~\ref{fig:diamonds-mode}. 
By contrast, the $w_{1+\infty}$ is an additional infinite symmetry enhancement that corresponds to higher powers of the energy and negative integer $\Delta$.
The wedge subalgebra, shaded in blue, is where we are able to realize a pure-matter representation of the celestial symmetries.\footnote{In a language akin to the soft physics Ward identities of~\cite{Strominger:2017zoo}, and up to subtleties about being globally well-defined on the celestial sphere~\cite{Pate:2019lpp,Himwich:2021dau,Donnay:2022sdg,Freidel:2022skz}, the soft charges (the appropriate smearings of $Q^1$ in~\eqref{Qk}) vanish in the wedge truncation.}
Just as the subalgebra of BMS that is unbroken by a choice of vacuum is Poincaré, we see that the intersection of the wedge and BMS is the chiral half of Poincaré that is consistent with the ansatz (\ref{equ:general-OPE}) for the anti-holomorphic OPE, namely $\{ P_{-\frac{1}{2},-\frac{1}{2}},\, P_{+\frac{1}{2},-\frac{1}{2}},\, L_{-1},\, L_0,\, L_1\}$\footnote{Here $P_{-\frac{1}{2},-\frac{1}{2}}$ and $P_{+\frac{1}{2},-\frac{1}{2}}$ are two global translations and $\{L_{-1},\, L_0,\, L_1\}$ are the global $SL(2)_L$ generators. For their actions on a primary, see (\ref{equ:W0m-Pm-action}) and (\ref{equ:W1m-Lm-action}).}.

This paper is organized as follows. In section~\ref{sec:wedge}, we start by recalling the celestial OPEs and the chiral celestial symmetry algebras. We then construct a phase space representation of the generators for the wedge subalgebra of celestial symmetries associated with generic spin-$s$ negative helicity modes. This representation is quadratic in the field operators and factorizes into radiative and matter contributions. In section~\ref{sec:charges}, we extend the phase space representation beyond the wedge truncation by introducing linear and higher-order operators.  Using a similar methodology, in section~\ref{sec:tilde-qs}, we construct a phase space realization of (a projection of) the positive helicity modes that transform under the chiral celestial symmetries. We close with a summary of what we have learned about detector operators for celestial symmetries and a discussion of future investigations in section~\ref{sec:conclusion}, followed by computational details in the appendix.

\section{Detector Operators for the Wedge Celestial Symmetries}\label{sec:wedge}

In celestial holography, scattering amplitudes in the bulk are dual to correlation functions of operators in a CFT living on the codimension-two celestial sphere. 
In the extrapolate dictionary, these celestial operators preparing the massless $in$/$out$ states can be thought of as living at the null conformal boundary. The soft limits of the external scattering state correspond to smearing the operators along the null time $u$: higher $u$-moments of the boundary fields map to various residues of the scaling dimension $\D$ of the corresponding celestial operator~\cite{Cheung:2016iub,Pate:2019mfs,Puhm:2019zbl,Adamo:2019ipt}. Meanwhile, the collinear limit of two external legs in momentum space corresponds to the coincidence limit on the celestial sphere.  

In this paper, we will focus on tree-level massless scattering. The leading OPE of two celestial primaries can be determined either by Mellin transforming the tree-level two-particle splitting functions in momentum space~\cite{Fan:2019emx,Pate:2019lpp} or more elegantly by demanding (chiral) Poincar\'e covariance and some assumptions on the analytic structure in $z_{ij}$ consistent with tree-level scattering to solve the OPE coefficients~\cite{Himwich:2021dau}. 
The celestial OPEs involving soft operators can be determined by taking the corresponding soft limits of the generic $\Delta$ OPEs.
To be more precise, let's consider the following tree-level massless 3-point collinear channel 
\begin{equation}
\begin{tikzpicture}[baseline={([yshift=-0.9ex]current bounding box.center)},scale=0.9]
        \draw[thick] (-1.2,0) node[above right]{$-J_p$} -- (0,0) node[below]{} -- (1.1,0) node[above left]{$J_p$};
        \draw[thick] (-1.5,0) -- (-2,1.73/2) node[left]{$J_1$};
        \draw[thick] (1.5,0) -- (2,1.73/2) node[right]{};
        \draw[thick] (1.5,0) -- (1.5+0.94,1.73/4) node[right]{};
        \draw[thick] (1.5,0) -- (1.5+0.94,-1.73/4) node[right]{};
        \draw[thick] (-1.5,0) -- (-2,-1.73/2) node[left]{$J_2$};
        \draw[thick] (1.5,0) -- (2,-1.73/2) node[right]{};
        \filldraw[thick,fill=white] (-1.5,0) circle (0.4) node{\tiny{$g_{12p}$}};
        \filldraw[thick,fill=white] (1.5,0) circle (0.4) node{};
        \filldraw[pattern=north west lines](1.5,0) circle (0.4);
\end{tikzpicture}
\end{equation}
where $g_{12p}$ denotes the bulk 3-point coupling constant for particles with helicity $J_1$, $J_2$, and $-J_p$. 
This diagram captures the leading singularity on the celestial sphere when particle 1 and 2 go collinear. To extract the celestial symmetry algebras we are interested in the complexified (anti-)holomorphic collinear limits. We will focus on the leading anti-holomorphic singularity ($1/\bz_{12}$) throughout this work.
The leading anti-holomorphic singularity in the celestial OPEs takes the following form~\cite{Himwich:2021dau}
\begin{equation}
\begin{split}
    {\cal O}^{\a}_{\D_1,J_1}(z_1,\bz_1)\,{\cal O}^{\b}_{\D_2,J_2}(z_2,\bz_2)~\sim&~ \sum_p\,\sum_{n=0}^{\infty}\, g^{\a\b\g}_{12p}\,\frac{z^{J_p-J_1-J_2-1}_{12}}{\bz_{12}}\,\frac{z_{12}^n}{n!}\\
    & B(\D_1-1-J_2+J_p+n,\D_2-1-J_1+J_p)\,\pa_{z_2}^n{\cal O}^{\g}_{\D_p,J_p}(z_2,\bz_2)~~, 
\end{split}
\label{equ:general-OPE}
\end{equation}
where $\D_p=\D_1+\D_2+J_p-J_1-J_2-2$.  To make our setup more general, we've allowed the operators to carry internal indices $\a$, $\b$, $\g$. For instance, if we take ${\cal O}^{\a}$ to be a gluon operator, $\a$ reduces to the color index $a$ and the three-point coupling $g^{abc}_{12p}$ is proportional to the structure constant $f^{abc}$.

In what follows we will focus on the case where $J_p=J_2$ and $J_1=-s_1$ ($s_1=|J_1|$). Namely, we are interested in taking a negative helicity operator ${\cal O}_1$ and extracting a set of bosonic currents that couple to a matter field ${\cal O}_2$ without changing its species or spin.\footnote{Note that for fermionic soft currents, the operator will be shifted within the supermultiplet. 
For further discussions about (conformally) soft photino, gaugino, and gravitino currents, see \cite{Dumitrescu:2015fej,Avery:2015iix,Fotopoulos:2020bqj,Pano:2021ewd}.} 
In this case (\ref{equ:general-OPE}) becomes
\begin{equation}
\begin{split}
        {\cal O}^{\a}_{\D_1,-s_1}(z_1,\bz_1)\,{\cal O}^{\b}_{\D_2,J_2}(z_2,\bz_2)~\sim~g^{\a\b\g}_{12p}\,\frac{z^{s_1-1}_{12}}{\bz_{12}}\,\sum_{n=0}^{\infty}\,B(\D_1-1+n,&2h_2-1+s_1)\,\frac{z_{12}^n}{n!}\\
        &\pa_{z_2}^n\,{\cal O}^{\g}_{\D_1+\D_2+s_1-2,J_2}(z_2,\bz_2)~~.
\end{split}
    \label{equ:OPE}
\end{equation}
Note that the Beta function has poles at certain integer values of $\D_1$. These poles exactly correspond to terms in the soft expansion of the scattering amplitudes. We define the soft mode of ${\cal O}^{\a}_{\D_1,-s_1}$ as follows
\begin{equation}
    H^{\a}_{k,s_1}(z,\bz) ~:=~ \Res_{\D_1=1-k}\,{\cal O}^{\a}_{\D_1,-s_1}(z,\bz)~~,
    \label{equ:def-Hks}
\end{equation}
where $k=0,1,2,\cdots$ is an non-negative integer. Before introducing the soft charges that form the celestial chiral symmetries, we first recall that a general conformal field with weights $(h,\bh)$ can be double mode expanded as follows
\begin{equation}
      \phi_{h,\bh}(z,\bz) ~=~ \sum_{m,n}\,z^{-h-m}\,\bz^{-\bh-n}\, \phi^{h,\bh}_{m,n}~~.
      \label{equ:CS-mode-exp}
\end{equation}
Then, one can identify a tower of higher spin-$k$ soft charges with weight $\Delta=1+s$ whose modes are related to those of the soft particles~\eqref{equ:def-Hks} as follows\footnote{Since we will use this notation in abundance, we emphasize here that: $m$ and $n$ distinguish the modes as in~\eqref{equ:CS-mode-exp} while, $s$ corresponds to the spin of the bulk gauge field, and the parameter $k$ is related to the boost weight of the corresponding radiative mode as in~\eqref{equ:def-Hks}. Note that the soft charges are related to the soft particles via light transforms~\cite{Strominger:2021mtt,Himwich:2021dau}, namely 
\begin{equation}
    \widehat{q}^{\a}_{k,s}~=~(-1)^{k+s+1}\,(k+s)!{\bf L}[{\cal O}_{1-k,-s}]~~.
    \label{equ:qhat-as-LT}
\end{equation}
This re-scaling and the one in (\ref{equ:hatq-mn}) is designed to make the soft current algebra take a simpler form~\cite{Strominger:2021mtt,Himwich:2021dau}. Under a Weyl reflection, the spins and weights get exchanged so that these charges indeed have spin $k$ and weight $\Delta=1+s$.
}
\begin{equation}
\widehat{q}^{\a,k,s}_{m,n}~=~ (-1)^{-\frac{k+s-1}{2}-m}\,\left(\frac{k+s-1}{2}-m\right)!\left(\frac{k+s-1}{2}+m\right)!\,H^{\a,k,s}_{m,n} ~~.
\label{equ:hatq-mn}
\end{equation}
The action of these modes on an operator was identified~\cite{Himwich:2021dau} by evaluating the anti-holomorphic commutator
\bea{l}\n
    \Big[\widehat{q}^{\a,k,s_1}_{m,n},\,{\cal O}^{\b}_{\D_2,J_2}(w,\bw)\Big]_{rad} ~:=~ \oint_{w}\,\frac{dz}{2\pi i}\,z^{\frac{s_1+k-1}{2}+m}\,\oint_{\bw}\,\frac{d\bz}{2\pi i}\,\bz^{\frac{s_1-k-1}{2}+n}\,\widehat{q}^{\a}_{k,s_1}(z,\bz)\,{\cal O}^{\b}_{\D_2,J_2}(w,\bw)~~    \sn\label{equ:action-QO}\\
    \qquad\qquad\qquad\qquad\qquad~~
    ~=~ {\cal D}_{m,n}^{k,s_1,\a\b\g}(h_2)\,{\cal O}^{\g}_{\D_2-k+s_1-1,J_2}(w,\bw)\sn \label{equ:[hatq,O]}
\eea
where we've defined the following differential operator 
\begin{equation}
    \begin{split}
        {\cal D}_{m,n}^{k,s_1,\a\b\g}(h_2) ~=~ i\,g_{12p}^{\a\b\g}\,\sum_{l=0}^k\begin{pmatrix}
      m+\frac{s_1+k-1}{2}\\
      l
      \end{pmatrix}(2h_2+&s_1-2)_l\frac{(s_1-1+k-l)!}{(k-l)!}\\
      & w^{\frac{s_1+k-1}{2}+m-l}\,\bw^{\frac{s_1-k-1}{2}+n}\,\pa_w^{k-l} ~~.
    \end{split}
    \label{equ:Differential-op}
\end{equation}
One can do an analogous analysis for the leading holomorphic sector. The holographic symmetry algebras of~\cite{Guevara:2021abz,Strominger:2021mtt} were identified by taking ${\cal O}^{\b}_{\D_2,J_2}$ as another soft charge.
For Yang-Mills theory and gravity, these take the form of the S algebra and $w_{1+\infty}$ respectively~\cite{Strominger:2021mtt}, which can be understood in terms of symmetries of self-dual gauge theories~\cite{Adamo:2021lrv,Ball:2021tmb,Costello:2022wso,Bu:2022iak}.

Now, inspired by~\cite{Freidel:2021ytz}, we want to identify charges $q^{\a,k,s_1}_{m,n}$\footnote{Here we drop the caret to distinguish the charges ${q}^{\a,k,s}_{m,n}$ that generate the celestial chiral symmetry algebra in phase space from the ones $\widehat{q}^{\a,k,s}_{m,n}$ that generate the transformations under the radial quantization bracket (\ref{equ:action-QO}).} that will generate the same transformations on phase space, where now instead of the radial quantization bracket~\eqref{equ:action-QO} we use the canonical commutation relations. Namely, $q^{\a,k,s_1}_{m,n}$ satisfies
\begin{equation}
    \Big[q^{\a,k,s_1}_{m,n},\,{\cal O}^{\b}_{\D_2,J_2}(z,\bz)\Big] ~:=~ {\cal D}_{m,n}^{k,s_1,\a\b\g}(h_2)\,{\cal O}^{\g}_{\D_2-k+s_1-1,J_2}(z,\bz)~~,
    \label{equ:quad-op-bracket}
\end{equation}
where ${\cal D}_{m,n}^{k,s_1,\a\b\g}(h_2)$ is the same as (\ref{equ:Differential-op}). Given the canonical commutation relations, one can solve equation (\ref{equ:quad-op-bracket}) for $q^{\a,k,s}_{m,n}$ in terms of fields in the phase space or the oscillators after the canonical quantization. Below we will show how this explicitly works and start with introducing fields of interest in the phase space. 

\paragraph{Fields in Phase Space}
Let $\Phi^{\a}_{\pm s}(u,z,\bz)$ denote the leading component in a large-$r$ expansion of the radiative fields and massless matter fields that they couple to. These fields have helicity $\pm s$. For example, 
\begin{equation}
    \begin{split}
       {\rm gravity}~:&~ \Phi_{+2}(u,z,\bz) ~=~ C_{zz}(u,z,\bz) ~~,~~ \Phi_{-2}(u,z,\bz) ~=~ C_{\bz\bz}(u,z,\bz) ~~,\\
       \text{Yang-Mills}~:&~ \Phi^a_{+1}(u,z,\bz) ~=~ A^a_{z}(u,z,\bz) ~~,~~ \Phi^a_{-1}(u,z,\bz) ~=~ A^a_{\bz}(u,z,\bz) ~~.
    \end{split}
\end{equation}
These fields admit the following free field canonical mode expansions 
\begin{equation}
    \begin{split}
        \Phi^{\a}_{+ s}(u,z,\bz) ~=&~ \frac{i}{(2\pi)^2}\,\int_0^{+\infty} d\o\,\Big[ a^{\dagger,\a}_{s}(\o,z,\bz)\,e^{i\o u} - b^{\a}_{s}(\o,z,\bz)\,e^{-i\o u}\Big] ~~,\\
        \Phi^{\a}_{- s}(u,z,\bz) ~=&~ \frac{i}{(2\pi)^2}\,\int_0^{+\infty} d\o\,\Big[ b^{\dagger,\a}_{s}(\o,z,\bz)\,e^{i\o u} - a^{\a}_{s}(\o,z,\bz)\,e^{-i\o u}\Big] ~~,
    \end{split}
    \label{equ:Phi-mode-exp}
\end{equation}
where the oscillators satisfy the following commutation relation
\begin{equation}
    \Big[ a^{\a}_s(\o,z,\bz), a^{\dagger,\b}_{s'}(\o',w,\bw)\Big] ~=~ (2\pi)^3\,\frac{2}{\o}\,\delta(\o-\o')\,\delta^{(2)}(z-w)\,\delta^{\a\b}\,\d_{ss'} ~~.
    \label{equ:aaJ-alg}
\end{equation}
The oscillators $b^{\a}_s$ and $b^{\dagger,\b}_{s'}$ satisfy the same commutation relation as (\ref{equ:aaJ-alg}), while $a^{\a}_s/a^{\dagger,\a}_{s}$ and $b^{\a}_s/b^{\dagger,\a}_{s}$ commute. 
The (outgoing) celestial conformal primary is defined as 
\begin{equation}
    \begin{split}
        {\cal O}^{\a}_{\D,\pm s}(z,\bz) ~:=~ \frac{\Gamma(\D-1)}{2}\,\int_{-\infty}^{+\infty} du\,u_+^{1-\D}\,\pa_u\,\Phi^{\a}_{\pm s}(u,z,\bz)~~,
    \end{split}
\end{equation}
where $u_+=u+i\,\epsilon$. In what follows we will suppress this subscript.

\paragraph{Identifying the Quadratic Charges}
Note that in the phase space construction, only the quadratic terms contribute to the LHS of (\ref{equ:quad-op-bracket}). Using the celestial sphere mode expansion of the charge 
\begin{equation}
 q^{\a}_{k,s}(z,\bz) ~=~ \sum_{m,n}\,z^{-\frac{1+s+k}{2}-m}\,\bz^{-\frac{1+s-k}{2}-n}\, q^{\a,k,s}_{m,n}
 \label{equ:charge-mode-exp}
 \end{equation}
and (\ref{equ:deltafun-mode-exp}-\ref{equ:GammafuncID}), the commutation relation~\eqref{equ:quad-op-bracket} between the quadratic operator and the conformal primary can be uplifted to the following bracket on the 4D phase space\footnote{This is the generalization for arbitrary $s_1$ and $J_2$ of (3.39) in~\cite{Hu:2022txx}.}
\begin{equation}
    \begin{split}
        \Big[q^{2,\a}_{k,s_1}(z_1,\bz_1),\,{\cal O}^{\b}_{\D_2,J_2}(z_2,\bz_2)\Big] ~=&~ i\,g^{\a\b\g}_{12p}\,\sum_{n=0}^{k}\,\frac{(-1)^{k-n}}{(k-n)!}\,(2h_2+s_1-2)_{k-n}\,\frac{(s_1-1+n)!}{n!}\\
        &\qquad\qquad\qquad\qquad \pa_{z_1}^{k-n}\d^{(2)}(z_{12})\,\pa_{z_2}^n{\cal O}^{\g}_{\D_2-k+s_1-1,J_2}(z_2,\bz_2) ~~,\\
    \end{split}
    \label{equ:[q2s,O]1}
\end{equation}
where the superscript ``$2$" on $q^{2,\a}_{k,s_1}$ means quadratic in fields/oscillators. Decomposing ${\cal O}^{\b}_{\D_2,J_2}$ into its creation and annihilation operator modes, (\ref{equ:[q2s,O]1}) reduces to the following commutation relations
\begin{equation}
    \begin{split}
        \Big[q^{2,\a}_{k,s_1}(z_1,\bz_1),\,a^{\dagger,\b}_{s_2}(\o,z_2,\bz_2)\Big] ~=&~ (-1)^{s_1}i\,g^{\a\g\b}_{12p}\,\sum_{n=0}^{k}\,\frac{(-1)^{k-n}}{(k-n)!}\,(\D_2+s_2+s_1-2)_{k-n}\\
        &\frac{(s_1-1+n)!}{n!}\, \pa_{z_1}^{k-n}\d^{(2)}(z_{12})\,(i\o)^{-k+s_1-1}\,\pa_{z_2}^na^{\dagger,\g}_{s_2}(\o,z_2,\bz_2) ~~,\\
        \Big[q^{2,\a}_{k,s_1}(z_1,\bz_1),\,a^{\b}_{s_2}(\o,z_2,\bz_2)\Big] ~=&~ i\,g^{\a\b\g}_{12p}\,\sum_{n=0}^{k}\,\frac{(-1)^{k-n}}{(k-n)!}\,(\D_2-s_2+s_1-2)_{k-n}\\
        & \frac{(s_1-1+n)!}{n!}\,\pa_{z_1}^{k-n}\d^{(2)}(z_{12})\,(-i\o)^{-k+s_1-1}\,\pa_{z_2}^n a^{\g}_{s_2}(\o,z_2,\bz_2) ~~.\\
    \end{split}
    \label{equ:[q2s,a]}
\end{equation}
Given that $q^{2,\a}_{k,s_1}$ is quadratic in the creation and annihilation operators, we can explicitly solve these two equations to find
\begin{equation}
    \begin{split}
        q^{2,\a}_{k,s_1(s_2)}(z,\bz) ~=&~ \frac{g^{\a\b\g}_{12p}}{2}\frac{1}{(2\pi)^3}\,\sum_{n=0}^{k}\,\frac{(-1)^{k-n}}{(k-n)!}\frac{(s_1-1+n)!}{n!}\,\int_0^{\infty}d\o_1\,\int_0^{\infty}d\o_2\,(-i\o_2)^{s_1-n}\\
        &\qquad \left(\frac{\o_1}{\o_2}\right)^{s_1-s_2}\,(-i\pa_{\o_1})^{k-n}\d(\o_1-\o_2)\,\pa_z^{k-n}\,\Big[ a^{\dagger,\b}_{s_2}(\o_1,z,\bz)\,\pa_z^n\,a^{\g}_{s_2}(\o_2,z,\bz)\Big] .
    \end{split}
    \label{equ:q2s-oscillator}
\end{equation}
Here the notation $q^{2,\a}_{k,s_1(s_2)}(z,\bz)$ means that we construct the charges that generate the spin-$s_1$ symmetry algebra in terms of spin-$s_2$ oscillators, generalizing how we added the matter contributions to the $Lw_{1+\infty}$ generators in~\cite{Hu:2022txx}. We show that (\ref{equ:q2s-oscillator}) indeed satisfies (\ref{equ:[q2s,a]}) in appendix \ref{appen:[q2s,O]}. 
It is also useful to write this expression in terms of the position space modes
\begin{equation}
   \begin{split}
      q^{2,\a}_{k,s_1(s_2)}(z,\bz) ~=&~ (-1)^{s_1-s_2}\frac{g^{\a\b\g}_{12p}}{2}\,\sum_{n=0}^{k}\,\frac{(-1)^{k-n}}{(k-n)!}\frac{(s_1-1+n)!}{n!}\,\int_{-\infty}^{\infty} du\,u^{k-n}\\
    &\qquad\qquad \pa_z^{k-n}\,\Big[ \pa_u^{s_1-s_2}\,\Phi^{\b}_{+s_2}(u,z,\bz)\,\pa_z^n\,\pa_u^{s_2-n}\Phi^{\g}_{-s_2}(u,z,\bz)\Big] ~~.
    \end{split}
\label{equ:q2s-field}
\end{equation}
Finally, by definition, $q^{2,\a,k,s_1}_{m,n}$ can be extracted by smearing the charge $q^{2,\a}_{k,s_1(s_2)}$ on the celestial sphere as follows
\begin{equation}
    \begin{split}
        q^{2,\a,k,s_1}_{m,n} ~=&~ \oint\,\frac{dz}{2\pi i}\,z^{\frac{s_1+k-1}{2}+m}\,\oint\,\frac{d\bz}{2\pi i}\,\bz^{-\frac{1-s_1+k}{2}+n}\,q^{2,\a}_{k,s_1(s_2)}(z,\bz) ~~.
    \end{split}
    \label{equ:qks,mn}
\end{equation}
Next, we will see that given the commutation relation (\ref{equ:quad-op-bracket}), one can derive the charge algebra via Jacobi identities.

\paragraph{Deriving Algebra via Jacobi Identities}
In~\cite{Himwich:2021dau} and also the recent work~\cite{Freidel:2023gue}, the authors show that for the gravitational case where $s_1=2$, the action (\ref{equ:quad-op-bracket}) guarantees that the quadratic operators form the wedge $w_{1+\infty}$. Similarly, one can show that for the Yang-Mills case where $s_1=1$, the quadratic operators form the wedge S algebra. We provide a brief proof for the wedge S algebra in appendix~\ref{appen:proof-YM}.
The idea is the following. Consider what happens when we apply the Jacobi identity after looking at two charges acting consecutively on one primary
\begin{equation}\resizebox{0.9\textwidth}{!}{$%
       \Big[q^{\a,k,s_1}_{m,n},\,  \Big[q^{\b,k',s_1}_{p,q},\,{\cal O}^{\g}_{\D_2,J_2}(z,\bz)\Big]\Big] - \Big[q^{\b,k',s_1}_{p,q},\,  \Big[ q^{\a,k,s_1}_{m,n},\,{\cal O}^{\g}_{\D_2,J_2}(z,\bz)\Big]\Big] 
      ~=~ \Big[\Big[ q^{\a,k,s_1}_{m,n},\,  q^{\b,k',s_1}_{p,q}\Big],\,{\cal O}^{\g}_{\D_2,J_2}(z,\bz)\Big] $}%
      \label{equ:JacobiID} 
\end{equation}
and for the last term, one can further evaluate it by plugging in the assumption that the charge algebra takes the following form\footnote{This assumption holds for gravity and Yang-Mills cases. We will comment on what happens for higher-spin $s_1>2$ in section \ref{sec:conclusion}.}
\begin{equation}
    \Big[ q^{\a,k,s_1}_{m,n},\,  q^{\b,k',s_1}_{p,q}\Big] ~=~ {\cal A}^{\a\b\d}_{m,n;p,q}(k,k';s_1)\,q^{\d,k+k'+1-s_1,s_1}_{m+p,n+q}~~.
    \label{equ:algebra-mode}
\end{equation}
Plugging in (\ref{equ:quad-op-bracket}) and (\ref{equ:algebra-mode}), equation (\ref{equ:JacobiID}) reduces to the following condition for the differential operator defined in (\ref{equ:Differential-op})
\begin{equation}
    \begin{split}
        {\cal D}_{p,q}^{k',s_1,\b\g\d}(h_2)\,{\cal D}_{m,n}^{k,s_1,\a\d\epsilon}\left(h_2-\frac{k'+1-s_1}{2}\right) -& {\cal D}_{m,n}^{k,s_1,\a\g\d}(h_2)\,{\cal D}_{p,q}^{k',s_1,\b\d\epsilon}\left(h_2-\frac{k+1-s_1}{2}\right) \\
        ~=&~ {\cal A}^{\a\b\d}_{m,n;p,q}(k,k';s_1)\,{\cal D}_{m+p,n+q}^{k+k'+1-s_1,s_1,\d\g\epsilon}(h_2) ~~.
    \end{split}
    \label{equ:cond-diff}
\end{equation}
Finally, plugging in (\ref{equ:Differential-op}), one can show that for gravity/Yang-Mills, the quadratic charges form the $w_{1+\infty}$/S algebra {\it only} within the wedge truncation. This means that we are able to represent this symmetry purely in the matter sector. In what follows, we will focus on the gravity and Yang-Mills examples and will come back to the higher-spin generalization in section~\ref{sec:conclusion}. 

\paragraph{Wedge Truncation} Since the wedge truncation plays an important role here, now let's review what it is.
Consider the mode expansion on the celestial sphere for soft particles defined in (\ref{equ:def-Hks}) as
\begin{equation}
    H^{\a}_{k,s}(z,\bz) ~=~ \sum_{m,n}\,z^{-\frac{1-s-k}{2}-m}\,\bz^{-\frac{1+s-k}{2}-n}\,H^{\a,k,s}_{m,n}~~.
    \label{equ:soft-mode-exp}
\end{equation}
The wedge truncation introduced in~\cite{Guevara:2021abz} restricts the range of $m$ to be $m\in[\frac{1-s-k}{2},\frac{s+k-1}{2}]$. This is equivalent to the following null state condition
\begin{equation}
    \pa_z^{k+s}\,H^{\a}_{k,s}(z,\bz) ~=~ 0 ~~.
    \label{equ:H-nullcond}
\end{equation}
This condition can be easily seen to be consistent with the leading anti-holomorphic sector in the soft theorems extracted from BCFW recursion relations~\cite{Guevara:2019ypd,Hu:2022bpa}, which takes the following form
\begin{equation}
    H^{\a}_{k,s}(z_1,\bz_1)\,{\cal O}_{h_2,\bh_2}(z_2,\bz_2) ~\sim~ \frac{1}{k!}\frac{z_{12}^{s-1}}{\bz_{12}}\,e^{(s-1)\pa_{\D_2}}\,\Big[\,e^{-\pa_{\D_2}}\,(z_{12}{\pa}_2-2h_2)  \,\Big]^k\,{\cal O}_{h_2,\bh_2}(z_2,\bz_2)~~.
\end{equation}
One can see that when acting with $\pa_{z_1}^{k+s}$ on both sides, the RHS vanishes up to contact terms.

\paragraph{Remarks} In this section, we construct the phase space representation of the quadratic operator mode (\ref{equ:qks,mn}) which forms the wedge subalgebra of celestial chiral symmetry algebra. Before extending this construction beyond the wedge truncation in section~\ref{sec:charges}, we end this section with several remarks.
\begin{itemize}
    \item Note that in (\ref{equ:q2s-field}), by choosing $\Phi^{\b}_{+s_2}$ and $\Phi^{\g}_{-s_2}$ to be massless matter fields, we are able to construct the representation of the wedge celestial chiral symmetry algebras completely in the matter sector. In appendix~\ref{appen:detector-op-wedge-examples}, we present explicit expressions for both gravity and Yang-Mills.
    \item In (\ref{equ:hatq-mn})  $\widehat{q}^{\a,k,s}_{m,n}$ was defined so that it forms celestial symmetry algebra via the radial quantization bracket (\ref{equ:action-QO}). Under the phase space canonical commutation bracket, the $\widehat{q}^{\a,k,s}_{m,n}$ transform in a  representation  of the charges $q^{2,\a,k,s}_{m,n}$. Namely, 
    \begin{equation}
    \Big[ q^{2,\a,k,s}_{m,n},\,  \widehat{q}^{\b,k',s}_{p,q}\Big] ~=~ {\cal A}^{\a\b\d}_{m,n;p,q}(k,k';s)\,\widehat{q}^{\d,k+k'+1-s,s}_{m+p,n+q}~~.
    \end{equation}
    \item Recall that in~\cite{Cordova:2018ygx,Hu:2022txx} we have seen that for the BMS subalgebra of $w_{1+\infty}$, its quadratic operator representation can be completely split into gravitational and matter sectors. Each of them forms the algebra and they are mutually commuting. In this section, we see this splitting also occurs for the wedge subalgebra of $w_{1+\infty}$.
    Schematically, we call quadratic operators hard charges, then we have the following factorization for both BMS subalgebra and wedge subalgebra
    \begin{equation}
     Q_{hard} ~=~ Q_{hard}^{grav} ~+~ Q_{hard}^{matter} ~~.
    \end{equation}
    \item As discussed in~\cite{Cordova:2018ygx,Hu:2022txx}, the matter sector representation is closely related to, and indeed generalizes, the ANEC and detector operators featured in conformal collider physics. Those operators are supported on a light ray (smeared along $u$), hence the name light-ray operators.
    In appendix~\ref{appen:detector-op-wedge-examples}, we explicitly show the relation between our quadratic operators formed in the matter sector and the light-ray operators for the gravity and Yang-Mills examples. Note that our quadratic operators that form the wedge subalgebra are smeared along $u,z,\bz$, thus they are supported on a light sheet instead of a light ray. We will have more discussions along these lines in section~\ref{sec:conclusion}.
\end{itemize}

\section{Celestial Symmetries of the Radiative Phase Space}\label{sec:charges}

As mentioned earlier, 
the goal of this section is two-fold: to extend results in section~\ref{sec:wedge} beyond the wedge truncation,
and also to extend the constructions presented in~\cite{Freidel:2021ytz,Hu:2022txx} where the $w_{1+\infty}$ symmetry charges are realized in the full radiative phase space. Namely, we will show how to systematically represent the generic spin-$s$ celestial chiral symmetry on the phase space.

The lesson we learned from the wedge truncation exercise in section~\ref{sec:wedge} is that the commutator between the charge mode and a primary operator (\ref{equ:quad-op-bracket}) guarantees that the charge mode forms the wedge subalgebra via the Jacobi identity. To realize (\ref{equ:quad-op-bracket}), we had to first uplift (\ref{equ:quad-op-bracket}) to a bracket on the 4D phase space (\ref{equ:[q2s,O]1}), and then project the quadratic operators to the wedge truncation for the algebra to close. To go beyond the wedge, we start with the solution of the quadratic operator obtained in (\ref{equ:q2s-oscillator}). Note that the Jacobi identity analysis used in both~\cite{Himwich:2021dau,Freidel:2023gue} and appendix~\ref{appen:Jacobi-wedge} only holds in the wedge truncation. From a simple example calculation beyond the wedge in appendix~\ref{appen:YMexample-failJacobi}, we can see that outside the wedge truncation, the commutator between quadratic operators does not close in general. Therefore, we next need to introduce linear and cubic operators, so that the bracket between linear and cubic operators cancels the anomalous terms in the quadratic-quadratic bracket. 

To motivate the linear operator and set up all the ingredients that we will need in this section and section~\ref{sec:tilde-qs}, we start with a discussion of the celestial diamond framework in section~\ref{sec:diamonds}. In section~\ref{sec:linear-truncation}, we define the linear operators and check the linear truncation of the algebra. Then the cubic operator is identified by forcing the quadratic truncation of the algebra close in section~\ref{sec:higher-order-op}.

\subsection{Celestial Diamonds}\label{sec:diamonds}

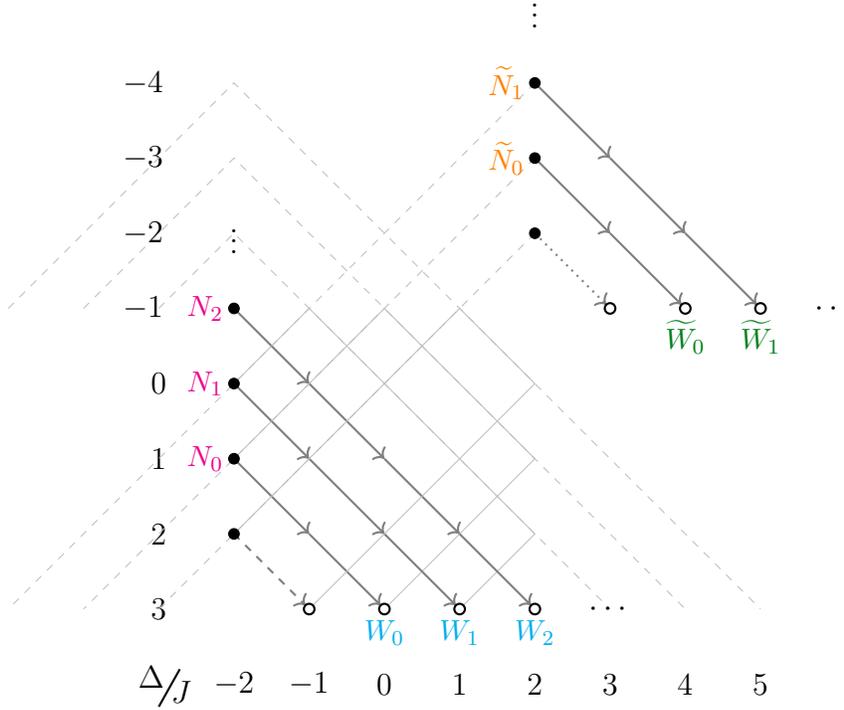
\begin{figure}[t]
    \centering
     \begin{tikzpicture}[scale=1]
\definecolor{darkgreen}{rgb}{.0, 0.5, .1};
\draw[lightgray] (0,3) -- (2,5) -- (4,3) -- (2,1);
\draw[lightgray] (0,4) -- (1,5) -- (4,2) -- (3,1);
\draw[lightgray,dashed] (-1,5) -- (0,6) --  (5,1);
\draw[lightgray,dashed] (-2,5) -- (0,7) --  (6,1);
\draw[->,gray,thick] (0,3) -- (1,2);
\draw[->,gray,thick]  (1,2)--(2-.05,1+.05);
\draw[->,gray,thick] (0,4) -- (1,3);
\draw[->,gray,thick] (1,3) -- (2,2);
\draw[->,gray,thick] (2,2) -- (3-.05,1+.05);
\draw[->,gray,thick] (0,5) -- (1,4);
\draw[->,gray,thick] (1,4) -- (2,3);
\draw[->,gray,thick] (2,3) -- (3,2);
\draw[->,gray,thick] (3,2) -- (4-.05,1+.05);
\node at (3,0) {$1$};
\node at (4,0) {$2$};
\node at (5,0) {$3$};
\node at (6,0) {$4$};
\node at (7,0) {$5$};
\draw[thick,fill=white] (2,1) circle (2pt) node[below]{\color{cyan}\small $W_0$} ;
\draw[thick,fill=white] (3,1) circle (2pt) node[below]{\color{cyan}\small $W_1$} ;
\node at (5,1) {$\cdots$};
\draw[thick,fill=white] (4,1) circle (2pt) node[below]{\color{cyan}\small $W_2$} ;
\node at (-.7,2.9-3)  {$J$};
\node at (-1.1,3.1-3) {$\tiny{\Delta}$};
\draw[thick] (-1,-.3) -- (-1+.28,.2);
\node at (0,6) {$\vdots$};
\node at (-1,1) {$3$};
\node at (-1,2) {$2$};
\node at (-1,3) {$1$~};
\node at (-1,4) {$0$~};
\node at (-1,5) {$-1$~~~~};
\node at (-1,6) {$-2$~~~~};
\node at (-1,7) {$-3$~~~~};
\node at (-1,8) {$-4$~~~~};
\node at (0,0) {$-2$};
\node at (1,0) {$-1$};
\node at (2,0) {$0$};
\node at (8,5) {$\cdots$};
\draw[->,gray,thick]  (0+4,3+4) -- (1+4,2+4);
\draw[->,gray,thick]  (1+4,2+4)--(2-.05+4,1+.05+4);
\draw[->,gray,thick] (0+4,4+4) -- (1+4,3+4);
\draw[->,gray,thick] (1+4,3+4) -- (2+4,2+4);
\draw[->,gray,thick] (2+4,2+4) -- (3-.05+4,1+.05+4);
\draw[lightgray,dashed] (4,7) --  (4-6,1);
\draw[lightgray,dashed] (4,8) --  (4-7,1);
\draw[lightgray,dashed] (-3,5) -- (0,8) --  (7,1);
\draw[->,gray,thick,dashed] (0,2) -- (1-.05,1+.05);
\draw[lightgray] (0,2) -- (3,5) -- (4,4) -- (1,1);
\draw[->,gray,thick,dotted] (4,6) -- (5-.05,5+.05);
\draw[lightgray,dashed] (4,6) -- (-1,1);
\filldraw[black] (0,3) circle (2pt) node[left]{\color{magenta}\small $N_0$} ;
\filldraw[black] (0,4) circle (2pt) node[left]{\color{magenta}\small $N_1$} ;
 \filldraw[black] (0,5) circle (2pt) node[left]{\color{magenta}\small $N_2$} ;
\filldraw[black] (4,6) circle (2pt);
\draw[thick,fill=white] (5,5) circle (2pt);
\filldraw[black] (4,7) circle (2pt) node[left]{\color{orange}\small $\widetilde{N}_{0}$} ;
\filldraw[black] (4,8) circle (2pt) node[left]{\color{orange}\small $\widetilde{N}_{1}$} ;
\node at (4,9) {$\vdots$};
\draw[thick,fill=white] (6,5) circle (2pt) node[below]{\color{darkgreen}\small $\widetilde{W}_0$} ;
\draw[thick,fill=white] (7,5) circle (2pt) node[below]{\color{darkgreen}\small $\widetilde{W}_1$} ;
\filldraw[black] (0,2) circle (2pt);
\draw[thick,fill=white] (1,1) circle (2pt);
\end{tikzpicture}
\caption{Celestial conformal dimensions and spins of the negative helicity soft modes giving rise to $w_{1+\infty}$ symmetry (magenta), the corresponding higher-spin symmetry generators (cyan) constructed from their primary descendants and modes in the matter sector, positive helicity soft graviton modes (orange), and the corresponding representation of the symmetry algebra (green) constructed from their primary descendants and modes in the matter sector. The spectra of these operators are related by 2D light transforms.  For reference, the `celestial diamonds' are superimposed in grey. Note that for the graviton primary operator, there is no pole at $\D=2$ and we use a dashed arrow to denote its primary descendant is zero in our analysis. For the positive helicity soft graviton modes, only those that couple to the negative helicity currents are shown. 
\label{fig:diamonds-s=2}}
\end{figure}

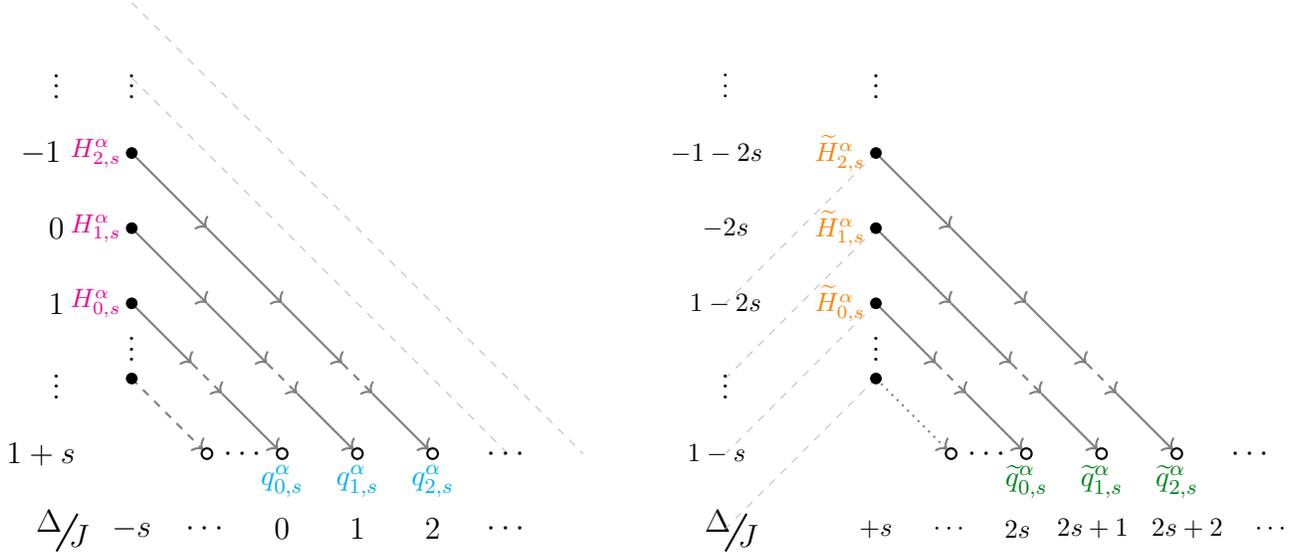
\begin{figure}[t]
\centering
\begin{minipage}[b]{0.45\textwidth}
\begin{tikzpicture}[scale=1]
\definecolor{darkgreen}{rgb}{.0, 0.5, .1};
\draw[lightgray,dashed]  (0,6) --  (5,1);
\draw[lightgray,dashed]  (0,7) --  (6,1);
\draw[->,gray,thick] (0,3) -- (0.8,3-0.8);
\draw[->,gray,thick,dashed] (0.8,3-0.8) -- (1.2,3-1.2);
\draw[->,gray,thick] (0.9+0.3,3-0.9-0.3) -- (2-.05,1+.05);
\draw[->,gray,thick] (0,4) -- (1,3);
\draw[->,gray,thick] (1,3) -- (1+0.8,3-0.8);
\draw[->,gray,thick,dashed] (1+0.8,3-0.8) -- (1+1.2,3-1.2);
\draw[->,gray,thick] (1+1.2,3-1.2) -- (3-.05,1+.05);
\draw[->,gray,thick] (0,5) -- (1,4);
\draw[->,gray,thick] (1,4) -- (2,3);
\draw[->,gray,thick] (2,3) -- (2+0.8,3-0.8);
\draw[->,gray,thick,dashed] (2+0.8,3-0.8) -- (2+1.2,3-1.2);
\draw[->,gray,thick] (2+1.2,3-1.2)-- (4-.05,1+.05);
\node at (3,0) {$1$};
\node at (4,0) {$2$};
\node at (-.7,2.9-3)  {$J$};
\node at (-1.1,3.1-3) {$\tiny{\Delta}$};
\draw[thick] (-1,-.3) -- (-1+.28,.2);
\filldraw[black] (0,3) circle (2pt) node[left]{\color{magenta}\footnotesize $H^{\a}_{0,s}$} ;
\filldraw[black] (0,4) circle (2pt) node[left]{\color{magenta}\footnotesize $H^{\a}_{1,s}$} ;
 \filldraw[black] (0,5) circle (2pt) node[left]{\color{magenta}\footnotesize $H^{\a}_{2,s}$} ;
\node at (-1,1) {$1+s$~~~~};
\node at (-1,2) {$\vdots$~};
\node at (-1,3) {$1$~};
\node at (-1,4) {$0$~};
\node at (-1,5) {$-1$~~~~};
\node at (-1,6) {$\vdots$};
\node at (0,6) {$\vdots$};
\node at (0,0) {$-s$};
\node at (1,0) {$\cdots$};
\node at (5,0) {$\cdots$};
\node at (2,0) {$0$};
\draw[->,gray,thick,dashed] (0,2) -- (1-.05,1+.05);
\node at (0,2.5) {$\vdots$};
\node at (1.5,1) {$\cdots$};
\draw[thick,fill=white] (2,1) circle (2pt) node[below]{\color{cyan}\small $q^{\a}_{0,s}$} ;
\draw[thick,fill=white] (3,1) circle (2pt) node[below]{\color{cyan}\small $q^{\a}_{1,s}$} ;
\node at (5,1) {$\cdots$};
\draw[thick,fill=white] (4,1) circle (2pt) node[below]{\color{cyan}\small $q^{\a}_{2,s}$} ;
\filldraw[black] (0,2) circle (2pt);
\draw[thick,fill=white] (1,1) circle (2pt);
\end{tikzpicture}
\end{minipage}\qquad
\begin{minipage}[b]{0.45\textwidth}
 \begin{tikzpicture}[scale=1]
\definecolor{darkgreen}{rgb}{.0, 0.5, .1};
\draw[->,gray,thick] (0,3) -- (0.8,3-0.8);
\draw[->,gray,thick,dashed] (0.8,3-0.8) -- (1.2,3-1.2);
\draw[->,gray,thick] (0.9+0.3,3-0.9-0.3) -- (2-.05,1+.05);
\draw[->,gray,thick] (0,4) -- (1,3);
\draw[->,gray,thick] (1,3) -- (1+0.8,3-0.8);
\draw[->,gray,thick,dashed] (1+0.8,3-0.8) -- (1+1.2,3-1.2);
\draw[->,gray,thick] (1+1.2,3-1.2) -- (3-.05,1+.05);
\draw[->,gray,thick] (0,5) -- (1,4);
\draw[->,gray,thick] (1,4) -- (2,3);
\draw[->,gray,thick] (2,3) -- (2+0.8,3-0.8);
\draw[->,gray,thick,dashed] (2+0.8,3-0.8) -- (2+1.2,3-1.2);
\draw[->,gray,thick] (2+1.2,3-1.2)-- (4-.05,1+.05);
\draw[thick,fill=white] (2,1) circle (2pt) node[below]{\color{darkgreen}\small $\widetilde{q}^{\a}_{0,s}$} ;
\draw[thick,fill=white] (3,1) circle (2pt) node[below]{\color{darkgreen}\small $\widetilde{q}^{\a}_{1,s}$} ;
\node at (5,1) {$\cdots$};
\draw[thick,fill=white] (4,1) circle (2pt) node[below]{\color{darkgreen}\small $\widetilde{q}^{\a}_{2,s}$} ;
\node at (-.7-1,2.9-3)  {$J$};
\node at (-1.1-1,3.1-3) {$\tiny{\Delta}$};
\draw[thick] (-1-1,-.3) -- (-1+.28-1,.2);
\node at (-2,1) {\footnotesize $1-s$~~~};
\node at (-2,2) {$\vdots$};
\node at (0,6) {$\vdots$};
\node at (-2,3) {\footnotesize $1-2s$};
\node at (-2,4) {\footnotesize $-2s$};
\node at (-2,5) {\footnotesize $-1-2s$~~~};
\node at (-2,6) {\footnotesize$\vdots$};
\node at (0,0) {\footnotesize $+s$};
\node at (1,0) {\footnotesize $\cdots$};
\node at (2,0) {\footnotesize $2s$~~~};
\node at (3,0) {\footnotesize $2s+1$~~~};
\node at (4,0) {~~\footnotesize $2s+2$};
\node at (5,0) {~~~~\footnotesize $\cdots$};
\draw[thick,fill=white] (1,1) circle (2pt);
\draw[->,gray,thick,dotted] (0,2) -- (1-.05,1+.05);
\node at (0,2.5) {$\vdots$};
\node at (1.5,1) {$\cdots$};
\draw[lightgray,dashed] (0,2) -- (-2,0);
\draw[lightgray,dashed] (0,3) -- (-2,1);
\draw[lightgray,dashed] (0,4) -- (-2,2);
\draw[lightgray,dashed] (0,5) -- (-2,3);
\filldraw[black] (0,3) circle (2pt) node[left]{\color{orange}\footnotesize $\widetilde{H}^{\a}_{0,s}$} ;
\filldraw[black] (0,4) circle (2pt) node[left]{\color{orange}\footnotesize $\widetilde{H}^{\a}_{1,s}$} ;
 \filldraw[black] (0,5) circle (2pt) node[left]{\color{orange}\footnotesize $\widetilde{H}^{\a}_{2,s}$} ;
\filldraw[black] (0,2) circle (2pt);
\end{tikzpicture}
\end{minipage} 
\caption{The generalization of the celestial diamonds in Fig. \ref{fig:diamonds-s=2} for radiative soft modes of spin-$s$, forming celestial chiral symmetry algebras (left) and representations thereof (right). 
Again, in the left diagram, the dashed arrow is used for modes that are zero in our analysis. In the right diagram, only modes that will contribute to the representation construction are labeled.  
\label{fig:diamonds} }
\end{figure}

The celestial diamond framework of~\cite{Pasterski:2021fjn,Pasterski:2021dqe} provides a nice way to organize the conformal multiplets: the symmetry charges reside at the bottom of the memory diamonds with $\Delta=1+s$ and are the primary ($L_{-1}$) descendants of radiative $J=- s$ conformally soft modes. The relevant celestial diamonds for $s=2$ are illustrated in Fig. \ref{fig:diamonds-s=2} and the generalizations to arbitrary spin-$s$ are in Fig. \ref{fig:diamonds}. Below we first explain how these two figures are obtained in section~\ref{sec:diamonds}. 

As a preliminary step, we will specify the conformal weights of all the operators of interest and explain our notation. First, the weights of the negative helicity conformally soft modes are given in (\ref{equ:def-Hks}). Their complex conjugates $\bar{H}^{\a}_{k,s}$ are radiative modes with positive helicity ($J=+s$).  We are interested in the $SL(2)_L$ primary descendants which occur when $h=\frac{1}{2}(\Delta+J)=\frac{1}{2}(1-n)$ for $n=0,1,2,..$.  In particular
\begin{center}
\begin{tabular}{cl}
   $J=-s$: &  $H^{\a}_{k,s}$ will have an $L_{-1}$ primary descendant at level $n={k+s}$, \\
   $J=+s$: & ${\bar H}^{\a}_{k,s}$ will have an $L_{-1}$ primary descendant at level $n={k-s}$. 
\end{tabular}
\end{center}
The conformal dimensions of these primary descendants are related to the ones of conformally soft modes by a Weyl reflection. As such the $SL(2)_L$ primary descendants of the $J=-s$ modes are at $\Delta'=1+s$ with spins $J'=-s+1,-s+2,...$ while the $SL(2)_L$ primary descendants of the $J=+s$ modes are at $\Delta'=1-s$ with spins $J'=s+1,s+2,...$~. The tree level amplitude is only expected to have poles starting at $\Delta=1$ (as can be seen in the beta function~\eqref{equ:OPE}) so for spin $s\ge 2$ we expect the $J'=-s+1,...,-1$ modes to vanish, hence the dashed arrow at the bottom left of figures~\ref{fig:diamonds-s=2} and \ref{fig:diamonds}, giving a tower of higher spin currents with spectrum $\Delta'=1+s$ and $J'=0,1,2...$~. These are illustrated by the cyan modes in figures~\ref{fig:diamonds-s=2} and \ref{fig:diamonds} which form a generic spin-$s$ symmetry algebra generalizing the $w_{1+\infty}$ of gravity. Although the anti-holomorphic collinear limits of positive helicity soft modes do not form a current algebra, they can be used to construct a representation of the aforementioned symmetry algebra. Starting from (\ref{equ:OPE}) we see that the ${\cal O}^{\a}_{\D_1,-s}{\cal O}^{\b}_{\D_2,+s}\sim {\cal O}^{\g}_{\D_p,+s}$ OPE only has poles starting at $\D_2=1-2s$. We thus get a tower of modes with $\Delta=1-s$ and $J=2s,2s+1,...$ illustrated in green in figures~\ref{fig:diamonds-s=2} and \ref{fig:diamonds}. Now the grey dotted arrows landing at $J=s+1,s+2,...,2s-1$ correspond to modes that do not appear in these OPEs.  We will return to this construction in Section~\ref{sec:tilde-qs}. Because the anti-holomorphic collinear limits treat negative and positive helicity modes differently, it is convenient to use slightly different notation 
\be
\widetilde{H}^{\a}_{k,s}(z,\bz) ~=~ \bar{H}^{\a}_{k+2s,s}(z,\bz)~~.
\ee
The shifted indices will make our expressions cleaner below.

In equation (\ref{equ:weight-table-1}), we list the conformal weights of the negative helicity soft modes for the gluon $R^a_k$, graviton $N_k$, and general spin-$s$ field $H^{\a}_{k,s}$. For their corresponding symmetry generators, we will use $J_k^a$ to denote the S algebra generators, $W_k$ for $w_{1+\infty}$, and $q^{\a}_{k,s}$ for the generic case. 
\begin{equation}
    \begin{tabular}{c|c|c}
       negative helicity soft modes  &  $(h,\bh)$ & $(\D,J)$\\\hline\hline
     gluon: $R^a_k(z,\bz)$  & $\left(\frac{-k}{2},\frac{2-k}{2}\right)$ & $(1-k,-1)$ \\
    $J_k^a$ (primary descendants of $R^a_k$: $\pa_z^{k+1}R^a_k$)  & $\left(\frac{2+k}{2},\frac{2-k}{2}\right)$ & $(2,k)$ \\ \hline\hline
    graviton: $N_k(z,\bz)$ & $\left(\frac{-1-k}{2},\frac{3-k}{2}\right)$ & $(1-k,-2)$ \\
    $W_k$ (primary descendants of $N_k$: $\pa_z^{k+2}N_k$)  & $\left(\frac{3+k}{2},\frac{3-k}{2}\right)$ & $(3,k)$ \\ \hline\hline
    helicity-$s$: $H^{\a}_{k,s}(z,\bz)$ & $\left(\frac{1-s-k}{2},\frac{1+s-k}{2}\right)$ & $(1-k,-s)$ \\
    $q^{\a}_{k,s}(z,\bz)$ (primary descendants of $H^{\a}_{k,s}$: $\pa_z^{k+s}H^{\a}_{k,s}$)  & $\left(\frac{1+s+k}{2},\frac{1+s-k}{2}\right)$ & $(s+1,k)$ \\ \hline
    \end{tabular}
    \label{equ:weight-table-1}
\end{equation}
Meanwhile (\ref{equ:weight-table-2}) provides a summary of the positive helicity soft modes of interest.
\begin{equation}
    \begin{tabular}{c|c|c}
       positive helicity soft modes  &  $(h,\bh)$ & $(\D,J)$\\\hline\hline
     gluon: $\widetilde{R}^a_k(z,\bz)$  & $\left(\frac{-k}{2},\frac{-2-k}{2}\right)$ & $(-1-k,+1)$ \\
    $\widetilde{J}_k^a$ (primary descendants of $\widetilde{R}^a_k$: $\pa_z^{k+1}\widetilde{R}^a_k$)  & $\left(\frac{k+2}{2},\frac{-2-k}{2}\right)$ & $(0,k+2)$ \\ \hline\hline
    graviton: $\widetilde{N}_k(z,\bz)$ & $\left(\frac{-1-k}{2},\frac{-5-k}{2}\right)$ & $(-3-k,+2)$ \\
    $\widetilde{W}_k$ (primary descendants of $\widetilde{N}_k$: $\pa_z^{k+2}\widetilde{N}_k$)  & $\left(\frac{3+k}{2},\frac{-5-k}{2}\right)$ & $(-1,k+4)$ \\ \hline\hline
    general helicity $s$: $\widetilde{H}^{\a}_{k,s}(z,\bz)$ & $\left(\frac{1-s-k}{2},\frac{1-3s-k}{2}\right)$ & $(1-2s-k,+s)$ \\
    $\widetilde{q}^{\a}_{k,s}(z,\bz)$ (primary descendants of $\widetilde{H}^{\a}_{k,s}$: $\pa_z^{k+s}\widetilde{H}^{\a}_{k,s}$)  & $\left(\frac{1+s+k}{2},\frac{1-3s-k}{2}\right)$ & $(1-s,k+2s)$ \\ \hline
    \end{tabular}
    \label{equ:weight-table-2}
\end{equation}

\subsection{Linear Truncation of the Algebra}\label{sec:linear-truncation}

First recall that in the last section, we promoted the symmetry charge bracket to a bracket on the 4D phase space and obtained the quadratic operator in terms of creation and annihilation oscillators in (\ref{equ:q2s-oscillator}) and field variables in (\ref{equ:q2s-field}). 
This form will let us make contact with the detector operators ${\cal E}_J$ in section~\ref{sec:conclusion}.

\subsubsection*{Linear Operators} 
As mentioned above, the quadratic-quadratic bracket on phase space doesn't close in general outside the wedge, therefore linear, cubic, and higher-order operators are necessary. Starting from the fact that inside the wedge truncation, the quadratic operators do satisfy the algebra, we can consider the case where the linear operator vanishes inside the wedge. 
The primary descendant of the soft mode is then a natural candidate for the linear term. First, as noted in our discussion of the celestial diamonds above, it has the correct spectrum; second, it vanishes in the wedge truncation.
We will thus take
\begin{equation}
    q^{1,\a}_{k,s}(z,\bz) ~=~ \pa_z^{k+s}\,H^{\a}_{k,s}(z,\bz)~~,
    \label{equ:q1k-def}
\end{equation}
which have weights $(\D,J)=(s+1,k)$. 
These linear charges can be expressed in terms of either field variables or the creation and annihilation operators as follows
\bea{l}\n \label{equ:q1s}
q^{1,\a}_{k,s}(z,\bz) ~=~  \frac{1}{2}\,\frac{(-1)^k}{k!}\,\int_{-\infty}^{+\infty} du\,u^k\,\pa_z^{k+s}\,\pa_u\,\Phi^{\a}_{-s}(u,z,\bz)~~,\sn
    \label{equ:q1s-fields}\\
   \qquad\quad ~=~ -\frac{1}{2}\,\frac{(-1)^k}{k!}\,\frac{i^k}{(2\pi)}\,\int_0^{\infty} d\o\,\o\,\pa_{\o}^{k}\d(\o)\,\pa_z^{k+s}\,\Big[ (-1)^k b^{\dagger,\a}_{s}(\o,z,\bz)  + a_s^{\a}(\o,z,\bz) \Big]~~.
     \sn\label{equ:q1s-oscillator}
\eea
In what follows we will use the Yang-Mills example as a point of reference.
Taking $s_1=s_2=1$, letting $\a$ be the color index $a$, and setting the coupling constant as $g^{abc}_{12p}=f^{abc}$ we have
\begin{equation}\label{Jlinear}
    \begin{split}
        J^{1,a}_{k}(z,\bz) ~=&~ \frac{1}{2}\,\frac{(-1)^k}{k!}\,\int_{-\infty}^{+\infty} du\,u^k\,\pa_u\,\pa_z^{k+1}\,A^a_{\bz}(u,z,\bz) ~~,\\
        ~=&~ -\frac{1}{2}\,\frac{(-1)^k}{k!}\,\frac{i^k}{(2\pi)}\,\int_0^{\infty} d\o\,\o\,\pa_{\o}^{k}\d(\o)\,\pa_z^{k+1}\,\Big[ (-1)^k b^{\dagger,a}_{1}(\o,z,\bz)  + a_1^{a}(\o,z,\bz) \Big] ~~.
    \end{split}
\end{equation}

\subsubsection*{Linear Truncation of the Algebra}\label{sec:linear-truncation-alg}
Since we already have the linear (\ref{equ:q1s}) and quadratic (\ref{equ:q2s-oscillator}-\ref{equ:q2s-field}) contributions to our charge operators, we are ready to evaluate the linear truncation of the spin-$s$ algebra. It consists of the following two contributions
\begin{equation}
    \Big[q^{\a}_{k,s}(z_1,\bz_1),\,q^{\b}_{k',s}(z_2,\bz_2)\Big]^1 ~=~  \Big[q^{2,\a}_{k,s}(z_1,\bz_1),\,q^{1,\b}_{k',s}(z_2,\bz_2)\Big] ~+~  \Big[q^{1,\a}_{k,s}(z_1,\bz_1),\,q^{2,\b}_{k',s}(z_2,\bz_2)\Big]~~ ,
    \label{equ:linear-truncation-split}
\end{equation}
where we've used the fact that the matter sector doesn't contribute to the linear truncation to take $s_1=s_2=s$. In appendix~\ref{appen:[q2k,q1k']} we show that the first term on the RHS of~\eqref{equ:linear-truncation-split} evaluates to
\begin{equation}
    \begin{split}
       \Big[q^{2,\a}_{k,s}(z_1,\bz_1),\,q^{1,\b}_{k',s}(z_2,\bz_2)\Big] 
        ~=&~ i\,g^{\a\b\g}_{12p}\,\sum_{p=0}^{k+k'+s}\,C^{(s)}(k,k';p)\, \pa_{z_1}^{p}\d^{(2)}(z_{12})\,\pa_{z_2}^{s-1-p}\,q^{1,\g}_{k+k'+1-s,s}(z_2,\bz_2)~~,
    \end{split}
    \label{equ:[q2k,q1k']}
\end{equation}
where the coefficient $C^{(s)}(k,k';p)$ is defined as
\begin{equation}
    C^{(s)}(k,k';p) ~:=~ \sum_{n=max[0,p-s-k']}^{min[k,p]}\,\frac{(s-1+k-n)!}{(k-n)!}\,\begin{pmatrix}
        k'+n\\
        k'
        \end{pmatrix}\,\begin{pmatrix}
        s+k'\\
        p-n
        \end{pmatrix}\,(-1)^{p-n} ~~.
        \label{equ:C(s)(k,k')-def}
\end{equation}
For the second term in (\ref{equ:linear-truncation-split}), we have
\begin{equation}
    \begin{split}
       \Big[q^{1,\a}_{k,s}(z_1,\bz_1),\,q^{2,\b}_{k',s}(z_2,\bz_2)\Big] ~=&~ -i\,g^{\b\a\g}_{12p}\,\sum_{p=0}^{k+k'+s}\,C^{(s)}(k',k;p)\,\sum_{n=0}^p(-1)^n\begin{pmatrix}
        p\\
        n
        \end{pmatrix}\\
        &\qquad\qquad\qquad\qquad \pa_{z_1}^{n}\d^{(2)}(z_{12})\,\pa_{z_2}^{s-1-n}\,q^{1,\g}_{k+k'+1-s,s}(z_2,\bz_2) ~~.
    \end{split}
    \label{equ:[q1k,q2k']}
\end{equation}
Note that while the left hand sides of (\ref{equ:[q2k,q1k']}) and (\ref{equ:[q1k,q2k']}) are related by a permutation of the indices and labels, the difference between the expressions on the RHS arises from manipulating the expression so that the charge is at $(z_2,\bz_2)$.

\paragraph{Yang-Mills Example}\label{sec:Kac-Moody-check}
As an explicit example, let us focus on the $s=1$ case allowing for a non-Abelian gauge symmetry. The linear operators are given in~\eqref{Jlinear}. Meanwhile, the quadratic S-algebra charges $J_{k,gluon}^{2,a}(z,\bz)$ are given by
\begin{equation}
    \begin{split}
        J^{2,a}_{k,gluon}(z,\bz) ~=&~ \frac{f^{abc}}{2}\,\sum_{l=0}^k\,\frac{(-1)^{k-l}}{(k-l)!}\,\int_{-\infty}^{\infty}du\,u^{k-l}\,\pa_{z}^{k-l}\Big[\, A^{b}_z(u,z,\bz)\,\pa_z^l\pa_u^{1-l} A^c_{\bz}(u,z,\bz)\,\Big] ~~.
    \end{split}
    \label{equ:J2-gluon}
\end{equation}
Here $f^{abc}$ is the structure constant of the gauge group. To consider matter fields in an arbitrary representation of the gauge group we leave $s_2$ unconstrained and take the coupling constant to $g^{a\b\g}_{12p}=i[T^a]_{\b\g}$ giving
\begin{equation}
        J^{2,a}_{k,matter}(z,\bz) ~=~\frac{(-1)^{1-s_2}i\,[T^a]_{\b\g}}{2}\,\sum_{l=0}^k\,\frac{(-1)^{k-l}}{(k-l)!}\,\int_{-\infty}^{\infty}du\,u^{k-l}\,\pa_{z}^{k-l}\Big[\, \big(\pa_u^{1-s_2}\Phi^{\b}_{s_2}\big)\,\pa_z^l\pa_u^{s_2-l} {\Phi}^{\g}_{-s_2}\,\Big]
\end{equation}
as our quadratic matter current that reproduces the coupling of the S-algebra to matter fields, i.e.~\eqref{equ:[q2s,O]1} for $s_1=1$.
Now let us turn to the algebra amongst the $s=1$ modes. For $s=1$ the coefficients~\eqref{equ:C(s)(k,k')-def} simplify to
\begin{equation}
    \begin{split}
       C^{(1)}(k,k';p)  ~=&~ \begin{cases}
        p\le k: ~~ \frac{(-1)^p}{p!\Gamma(1-p)}~~~ \text{only nonzero when } p=0 ~~,\\
        p\ge k+1: ~~ (-1)^{p+k}\,\begin{pmatrix}
        p-1\\
        k
        \end{pmatrix}\,\begin{pmatrix}
        k+k'+1\\
        p
        \end{pmatrix} ~~.
        \end{cases}
    \end{split}
    \label{equ:C(1)}
\end{equation}
Therefore, we have
\begin{equation}
    \begin{split}
        & \Big[J^{2,a}_{k}(z_1,\bz_1),\,J^{1,b}_{k'}(z_2,\bz_2)\Big] ~=~ if^{abc}\,\delta^{(2)}(z_{12})\,J^{1,c}_{k+k'}(z_2,\bz_2) \\
        &~+~ if^{abc}\,\sum_{p=k+1}^{k+k'+1}\,(-1)^{p+k}\,\begin{pmatrix}
        p-1\\
        k
        \end{pmatrix}\,\begin{pmatrix}
        k+k'+1\\
        p
        \end{pmatrix}\,\pa_{z_1}^{p}\delta^{(2)}(z_{12})\,\pa^{-p}_{z_2}\,J^{1,c}_{k+k'}(z_2,\bz_2) ~~.\\
    \end{split}
    \label{equ:[J2a,J1b]-final}
\end{equation}
Similarly, as shown in appendix~\ref{appen:tildeC}, the second contribution to the linear truncation of the algebra can be simplified as follows
\begin{equation}
    \begin{split}
         \Big[J^{1,a}_{k}(z_1,\bz_1),\,J^{2,b}_{k'}(z_2,\bz_2)\Big] 
        ~=~ if^{abc}\,\sum_{n=k+1}^{k+k'+1}\,(-1)^{k+n+1}\,& \begin{pmatrix}
        k+k'+1\\
        n
        \end{pmatrix}\,\begin{pmatrix}
        n-1\\
        k
        \end{pmatrix}\\
        & \pa_{z_1}^{n}\delta^{(2)}(z_{12})\,\pa^{-n}_{z_2}\,J^{1,c}_{k+k'}(z_2,\bz_2) ~~. \\
    \end{split}
    \label{equ:[J1a,J2b]-final}
\end{equation}
Summing (\ref{equ:[J2a,J1b]-final}) and (\ref{equ:[J1a,J2b]-final}), we obtain the linear truncation of the S-algebra 
\begin{equation}\label{Jtrucate}
    \Big[J^{a}_{k}(z_1,\bz_1),\,J^{b}_{k'}(z_2,\bz_2)\Big]^1 ~=~ if^{abc}\,\delta^{(2)}(z_{12})\,J^{1,c}_{k+k'}(z_2,\bz_2) ~~
\end{equation}
as desired.
Finally, let us focus on the Kac-Moody subalgebra corresponding to the leading spin-0 ($k=0$) charges. The contribution from the
matter sector 
\begin{equation}
    J^{2,a}_{0,matter}(z,\bz) ~=~ \frac{i[T^a]_{\b\g}}{4}\,\int_{-\infty}^{\infty}du\, \Phi^{\b}_{+s}(u,z,\bz)\,\overset{\leftrightarrow}{\pa}_u \Phi^{\g}_{-s}(u,z,\bz) ~=~ \int_{-\infty}^{\infty}du\,j_{u,matter}^a(u,z,\bz)~~
    \label{equ:charge-flux-op-matter}
\end{equation}
is precisely the non-Abelian analog of the charge flux operator studied in \cite{Hofman:2008ar,Cordova:2018ygx}. Computing the commutator, we see that
\begin{equation}
    \begin{split}
        \Big[ J_{0,matter}^{2,a}(z,\bz), J_{0,matter}^{2,b}(w,\bw)\Big] 
        ~=&~\d^{(2)}(z-w)\,\Big\{ [T^a]_{\b\d}\,[T^b]_{\d\g} -  [T^a]_{\a\g}\,[T^b]_{\b\a} \Big\} \\ 
        & \qquad \frac{1}{2}\frac{1}{(2\pi)^3}\,\int_0^{\infty}d\o\,\o\,a^{\dagger,\b}_s(\o,w,\bw) a^{\g}_s(\o,w,\bw) \\
        ~=&~ i\,f^{abc}\,\d^{(2)}(z-w) \,J_{0,matter}^{2,c}(w,\bw)~~.
    \end{split}
    \label{equ:q20-Kac-Moody}
\end{equation}
Namely, the $k=0$ matter charges obey a representation of the Kac-Moody symmetry, even when we decouple the gauge field. This was observed to hold for these parallel light-ray operators in any unitary 4D CFT by Cordova and Shao~\cite{Cordova:2018ygx} and is the gauge theory analog of the fact that the matter stress tensor can be used to construct a representation of the BMS group.
Setting $s=1$ and $[T^a]_{\b\g}=-if^{abc}$ in the above calculation, one can see that the gluon sector commutator takes the same form. Together with~\eqref{Jtrucate}, we see that the gluon sector obeys an independent representation of the Kac-Moody symmetry.

\subsection{Higher-order Operators}\label{sec:higher-order-op}

From our discussions in~\cite{Hu:2022txx} we saw that the factorization between the gravitational and matter sectors does not continue beyond the BMS subalgebra.  Indeed, for $k\ge 2$ cubic and higher order operators are needed to make the algebra close. We will now turn to the S-algebra case. 

\paragraph{\texorpdfstring{Constructing the $k=1$ Charge in Yang-Mills}{Construction of J3a}}
Unlike the $k=0$ counterpart, for $k=1$ the commutator of the quadratic charges receives contributions from both gluon and matter sectors and does not close on their own,
\begin{equation}
    \begin{split}
        &\Big[ J_{0}^{2,a}(z,\bz), J_{1}^{2,b}(w,\bw)\Big]_{gluon/matter}  ~=~ i\,f^{abc}\,\d^{(2)}(z-w)\,J_{1,{gluon/matter}}^{2,c}(w,\bw)\\
        &\qquad ~+~\frac{i}{2}\frac{1}{(2\pi)^3}\,\pa_w\d^{(2)}(z-w)\,\Big\{\,-\, [T^b]_{\b\d}[T^a]_{\d\g}\,\int_0^{\infty}d\o\,{\rm a}^{\dagger,\b}_J(\o,w){\rm a}^{\g}_J(\o,w)\\
        &\qquad\qquad~+~ i\,f^{abc}\,[T^c]_{\b\g}\,\int_0^{\infty}d\o_2\,\int_0^{\infty}d\o_3\,\o_3\,\pa_{\o_2}\d(\o_2-\o_3)\,{\rm a}^{\dagger,\b}_J(\o_2,w){\rm a}^{\g}_J(\o_3,w) \,\Big\}~~,
    \end{split}
    \label{equ:[J0,J1]}
\end{equation}
where to be general, the quadratic operator is constructed by spin-$J$ fields. The gluon sector and the matter sector shares the same form as (\ref{equ:[J0,J1]}), where for the gluon sector we set $[T^a]_{\b\g}=-if^{abc}$, $\b=b$, $\g=c$, and $J=1$.
The $J_{1}^{3,a}(z,\bz)$ operator that resolves this can be solved via the quadratic truncation of the algebra with $k=0$, $k'=1$. 
Namely,
\begin{equation}\label{truncate}
    \begin{split}
         &i\,f^{abc}\,\d^{(2)}(z-w)\,J^{2,c}_{1,{gluon/matter}}(w,\bw) ~=~ \Big[ J_0^{a}(z,\bz), J_1^{b}(w,\bw)\Big]^2_{gluon/matter} \\
        &~=~ \Big[ J_0^{1,a}(z,\bz), J_1^{3,b}(w,\bw)\Big]_{gluon/matter} ~+~ \Big[ J_0^{2,a}(z,\bz), J_1^{2,b}(w,\bw)\Big]_{gluon/matter} ~~.
    \end{split}
\end{equation}
Plugging in the gluon and matter modes for $J_{k\le1}^{n\le2,a}$ it is straightforward to check that~\eqref{truncate} is satisfied for
\begin{equation}
    \begin{split}
         J_{1,gluon}^{3,b}(w,\bw) ~=&~ \frac{1}{2}\,f^{bcd}\,f^{def}\,\int_{-\infty}^{\infty}du\,A^c_z\,\pa_u^{-1}\,\Big(\,A^e_z\,\pa_u\,{A}^f_{\bz} \,\Big) ~~,
    \end{split}
\end{equation}
and
\begin{equation}
    \begin{split}
        J_{1,matter}^{3,b}(w,\bw) ~=~ \frac{(-1)^{s+1}}{2}\,\Bigg[\,&i\,f^{bcd}\,[T^d]_{\b\g}\,\int_{-\infty}^{\infty}du\,A^c_z\,\pa_u^{-1}\,\Big(\,\pa_u^{1-J}\Phi_J^{\b}\,\pa^J_u\,{\Phi}_{-J}^{\g} \,\Big) \\
    &+~[T^b]_{\b\d}\,[T^e]_{\d\g}\,\int_{-\infty}^{\infty}du\,\pa_u^{1-J}\Phi_J^{\b}\,\pa_u^{-1}\,\Big(\,A^e_z\,\pa^s_u\,{\Phi}_{-J}^{\g} \,\Big)\,\Bigg] ~~.
    \end{split}
\end{equation}
As discussed in appendix~\ref{appen:proof-YM}, the $k\le 1$ terms are enough to generate the rest of the S-algebra. Similar to the gravity case as in~\cite{Hu:2022txx}, cubic matter charges involve both gauge fields and matter fields. In other words, to obtain the matter representation we have to turn on the couplings to gauge fields, although one is able to completely turn off the matter sector and construct the pure radiative phase space representation.

\section{Representations from the Opposite Helicity Sector}\label{sec:tilde-qs}

When discussing the celestial chiral symmetry algebra and its generators in previous sections, we only considered a single helicity sector. Namely in (\ref{equ:general-OPE}), $J_1$ and $J_2$ have the same sign and in this work, we focus on the negative one. However, our discussion of the matter fields did not impose such a restriction. 
Indeed, we expect the opposite helicity modes to transform in a representation of the spin-$s$ celestial chiral symmetry algebra. We'll denote the representation encoded in the anti-holomorphic collinear singularities by $\widetilde{q}^{\a}_{k,s}(z,\bz)$ to distinguish these generators from the complex conjugates of our symmetry charges.
In summary, the anti-holomorphic collinear singularities imply the following structure of commutation relations
\begin{equation}
    \begin{split}
         \Big[ q^{\a}_{k,s}(z,\bz), q^{\b}_{k',s}(w,\bw) \Big] ~=&~ {\cal A}^{\a\b\g}(k,k';s)\,q^{\g}_{k+k'+1-s,s}(w,\bw) ~~,\\
         \Big[ q^{\a}_{k,s}(z,\bz), \widetilde{q}^{\b}_{k',s}(w,\bw) \Big] ~=&~ {\cal A}^{\a\b\g}(k,k';s)\,\widetilde{q}^{\g}_{k+k'+1-s,s}(w,\bw) ~~,\\
         \Big[ \widetilde{q}^{\a}_{k,s}(z,\bz), \widetilde{q}^{\b}_{k',s}(w,\bw) \Big] ~=&~ 0~~.
    \end{split}
    \label{equ:q-tildeq-alg}
\end{equation}
where the constants ${\cal A}^{\a\b\g}(k,k';s)$ in the first two lines are the same since they are coming from the same three-point vertex in the bulk.
Based on the celestial diamond discussion in section~\ref{sec:charges}, we define the linear operator $\widetilde{q}^{1,\a}_{k,s}$ as the ${\rm SL}(2)_L$ primary descendant at level $k+s$ of the positive helicity soft mode
\begin{equation}
   \widetilde{q}^{1,\a}_{k,s}(z,\bz) ~:=~  \pa_z^{k+s}\,\widetilde{H}^{\a}_{k,s}(z,\bz) ~~,
\end{equation}
where 
\begin{equation}
     \widetilde{H}^{\a}_{k,s}(z,\bz) ~=~  \frac{1}{2}\,\frac{(-1)^k}{(k+2s)!}\,\int_{-\infty}^{+\infty} du\,u^{k+2s}\,\pa_u\,\Phi^{\a}_{+s}(u,z,\bz)\\
\end{equation}
has $(\D,J)=(1-k-2s,+s)$ and $\widetilde{q}^{1,\a}_{k,s}$ has $(\D,J)=(1-s,k+2s)$. 
As we will now show, the bracket between $ q^{\a}_{k,s}(z,\bz)$ and $\widetilde{q}^{\b}_{k',s}(w,\bw)$ in the second line of (\ref{equ:q-tildeq-alg}) can be realized following a similar procedure to what we used in section~\ref{sec:charges} to reproduce the first line of (\ref{equ:q-tildeq-alg}). We will then get the third line for free, due to the fact that $\widetilde{q}^{\a}_{k,s}$ operators now only involve modes of the same helicity, and their bracket on phase space vanishes.

\paragraph{Linear truncation} First, following the similar computation as in appendix~\ref{appen:[q2k,q1k']}, we have
\begin{equation}
    \begin{split}
        \Big[ q^{2,\a}_{k,s}(z,\bz), \widetilde{H}^{\b}_{k',s}(w,\bw) \Big] ~=&~ i^{2s+1}\,g^{\a\g\b}_{12p}\,\sum_{n=0}^k\,\frac{(s-1+n)!}{n!}\,\begin{pmatrix}
    k+k'-n\\
    k'
    \end{pmatrix}\\
    &\qquad\qquad\qquad\qquad \,\pa_z^{k-n}\d^{(2)}(z-w)\pa_w^n\,\widetilde{H}^{\g}_{k+k'+1-s,s}(w,\bw)~~.
    \end{split}
    \label{equ:q2-tildeH}
\end{equation}
Forcing the linear truncation of (\ref{equ:q-tildeq-alg}) to work yields
\begin{equation}
\begin{split}
    \Big[ \widetilde{q}^{2,\a}_{k,s}(z,\bz), {H}^{\b}_{k',s}(w,\bw) \Big] ~=&~ i\,g_{12p}^{\a\b\g}\,\sum_{n=0}^k\,\frac{(s-1+n)!}{n!}\begin{pmatrix}
        k+k'-n\\
        k'
        \end{pmatrix}\\
        &\qquad\qquad\qquad\qquad \pa_z^{k-n}\d^{(2)}(z-w)\,\pa_w^n\,\widetilde{H}^{\g}_{k+k'+1-s,s}(w,\bw) ~~.
\end{split}
\label{equ:condition-tildeq}
\end{equation}
As shown in appendix~\ref{appen:check-tildeq}, the solution of (\ref{equ:condition-tildeq}) is given by
\begin{equation}
    \begin{split}
        \widetilde{q}^{2,\a}_{k,s}(z,\bz) ~=&~  \frac{1}{2}\frac{g^{\a\b\g}_{12p}}{2}\,\sum_{n=0}^{k}\,\frac{(-1)^{k-n}}{(k-n)!}\frac{(s_1-1+n)!}{n!}\,\int_{-\infty}^{\infty} du\,u^{k-n}\\
    &\qquad\qquad \pa_z^{k-n}\,\Big[ \,\Phi^{\b}_{+s}(u,z,\bz)\,\pa_z^n\,\pa_u^{-s-n}\Phi^{\g}_{+s}(u,z,\bz)\Big] ~~.
    \end{split}
    \label{equ:general-tilde-q}
\end{equation}
Then, with (\ref{equ:q2-tildeH}) and (\ref{equ:condition-tildeq}), one can follow the same calculation in section~\ref{sec:linear-truncation-alg} to show that the linear truncation of the second line in (\ref{equ:q-tildeq-alg}) holds.
As mentioned above, we indeed see that $\widetilde{q}^{\a}_{k,s}(z,\bz)$ commute with each other automatically because they only involve a single helicity mode.
For concreteness, we will now look at two important examples of interest: Yang-Mills and gravity. 

\paragraph{\texorpdfstring{$\widetilde{J}^a_k(z,\bz)$}{widetilde{J}ak} in Yang-Mills} The linear and quadratic charges  $\widetilde{J}^a_k(z,\bz)$ in pure Yang-Mills are given by
\begin{equation}
    \begin{aligned}
        \widetilde{J}^{1,a}_{k}(z,\bz) ~=&~ \frac{1}{2}\frac{(-1)^k}{(k+2)!}\,\int du\,u^{k+2}\,\pa^{k+1}_z\,\pa_u\,A_z^a(u,z,\bz)~~,\\
        \widetilde{J}^{2,a}_{k}(z,\bz) ~=&~ \frac{f^{abc}}{4}\,\sum_{l=0}^k\,\frac{(-1)^{k-l}}{(k-l)!}\,\int_{-\infty}^{\infty}du\,u^{k-l}\,\pa_{z}^{k-l}\Big[\, A^{b}_z(u,z,\bz)\,\pa_z^l\pa_u^{-1-l} A^c_{z}(u,z,\bz)\,\Big] ~~. \\  
    \end{aligned}
    \label{equ:tildeJ}
\end{equation}

\paragraph{\texorpdfstring{$\widetilde{W}_k(z,\bz)$}{widetilde{W}k} in Gravity} The linear and quadratic charges  $\widetilde{W}_k(z,\bz)$ in pure gravity are given by
\begin{equation}
    \begin{split}
        \widetilde{W}^1_k(z,\bz) ~=&~ \frac{1}{2}\frac{(-1)^{k}}{(k+4)!}\,\int du\,u^{k+4}\,\pa^{k+2}_z\,\pa_u\,C_{zz}(u,z,\bz) ~~,\\
        \widetilde{W}^2_k(z,\bz) ~=&~ \frac{1}{2}\frac{1}{4}\,\sum_{l=0}^k\,\frac{(-1)^{k-l}(l+1)}{(k-l)!}\,\int_{-\infty}^{\infty}du\,u^{k-l}\,\pa_{z}^{k-l}\Big[\, C_{zz}(u,z,\bz)\,\pa_z^l\pa_u^{-2-l} C_{zz}(u,z,\bz)\,\Big] ~~. \\ 
    \end{split}
\end{equation}

As in section~\ref{sec:charges} one can similarly add matter contributions in the quadratic operators and construct the higher-order operators following a similar procedure to section~\ref{sec:charges}. Namely, evaluating the quadratic or higher-order truncation of the bracket (\ref{equ:q-tildeq-alg}) yields higher-order operators for $\widetilde{q}^{\a}_{k,s}$. We leave an example of $\widetilde{J}^{3,a}_{1}$ for Yang-Mills in appendix~\ref{appen:example-Jtilde}.
One thing to keep in mind is that the linear operators $q^{1,\a}_{k,s}(z,\bz)$ and $\widetilde{q}^{1,\a}_{k,s}(z,\bz)$ are constructed by opposite helicity modes, therefore their commutator is no longer vanished. 
Namely, there is an additional central term
\begin{equation}
   \Big[ q^{\a}_{k,s}(z,\bz), \widetilde{q}^{\b}_{k',s}(w,\bw) \Big]^0 ~=~ \d^{\a\b}\, \widetilde{q}^0_{k+k'+1-s,s}\,\pa_w^{k+k'+2s}\d^{(2)}(z-w) ~~,
\end{equation}
where 
\begin{equation}
    \widetilde{q}^0_{k+k'+1-s,s} ~=~ \pi\,i^{k+k'}\,\Big[(-1)^{k+k'}-1\Big]\,\int_0^{\infty}\frac{d\o}{\o^{k+k'+2s-1}}\,\d(\o)\,\d(\o) ~~.
\end{equation}

\paragraph{Wedge Truncation} Given (\ref{equ:q-tildeq-alg}) one can implement celestial sphere mode expansion (\ref{equ:CS-mode-exp}) on both sides and obtain
\begin{equation}
    \begin{split}
        \Big[ q^{\a,k,s}_{m,n},\,  \widetilde{q}^{\b,k',s}_{p,q}\Big] ~=&~ {\cal A}^{\a\b\g}_{m,n;p,q}(k,k';s)\,\widetilde{q}^{\g,k+k'+1-s,s}_{m+p,n+q}~~.
    \end{split}
\end{equation}
Again, projecting the quadratic operator (\ref{equ:general-tilde-q}) to the wedge truncation yields the representation of the wedge subalgebra constructed from the opposite helicity sector
\begin{equation}
    \begin{split}
        \widetilde{q}^{2,\a,k,s}_{m,n} ~=&~ \oint\,\frac{dz}{2\pi i}\,z^{\frac{s+k-1}{2}+m}\,\oint\,\frac{d\bz}{2\pi i}\,\bz^{-\frac{1+3s+k}{2}+n}\,\widetilde{q}^{2,\a}_{k,s}(z,\bz) ~~,
    \end{split}
    \label{equ:tilde-qks,mn}
\end{equation}
where $m\in[\frac{1-s-k}{2},\frac{s+k-1}{2}]$.

\section{Discussion}\label{sec:conclusion}

The underlying motivation of this paper is to understand how to organize the radiative phase space in light of celestial symmetries.
What we explicitly computed are phase space realizations of the celestial symmetry generators for massless matter fields coupled to bulk gauge fields of arbitrary spin. In particular, we identified a pure matter realization of the wedge subalgebra, the general-spin analog of $w_{1+\infty}$ charges beyond the wedge, and light-ray operators that capture the opposite helicity modes they couple to.
We'll close by discussing various natural routes for further investigations.

\paragraph{Higher-spin Detector Operators}
In the above sections, whenever we wanted explicit examples, we focused on Yang-Mills (spin-1) and gravity (spin-2). However, the way that we defined detector operators in section~\ref{sec:wedge} allows for a generalization to higher spins. Let's first recap the setup for the detector operators for celestial chiral symmetries. 
\begin{enumerate}
    \item 
    First, we saw that chiral-Poincaré  -- a subgroup of the global Poincaré isometries that contains two translations and ${\rm SL}(2)_L$ --  covariance and the tree-level collinear limits fixed the celestial OPEs (\ref{equ:general-OPE}) for generic spins up to a coefficient fixed by the 3-point coupling constant in the bulk EFT.\footnote{One can also use  dimensional analysis and little group covariance to derive the 3-point amplitude.}
    \item Then, following the procedure in~\cite{Himwich:2021dau}, we  implemented a light transform (\ref{equ:qhat-as-LT}) that rescaled the terms appearing in the mode expansion of the OPE, and obtained the charge action (\ref{equ:[hatq,O]}) by computing the anti-holomorphic commutator.
    \item Finally, we promoted (\ref{equ:[hatq,O]}) to a canonical bracket on phase space and identified a representation of the charges. Namely, we showed that we could construct an operator $q^{2,\a}_{k,s_1(s_2)}(z,\bz)$ in (\ref{equ:q2s-oscillator}-\ref{equ:q2s-field}) that satisfies (\ref{equ:quad-op-bracket}) via the canonical quantization commutation relations (\ref{equ:aaJ-alg}). Here, $q^{2,\a}_{k,s_1(s_2)}(z,\bz)$ is quadratic in the spin-$s_2$ fields, has conformal scaling dimension $\D=1+s_1$ and 2D spin $J=k$, and is local on the celestial sphere but non-local in the $u$ direction.
\end{enumerate}
Now consider an $s_1=s>2$ gauge field coupled to a free complex scalar field $\Phi$ $(s_2=0)$. For simplicity, in what follows, we consider no color structure. The spin-zero charge corresponds to the higher-spin ANEC ${\cal E}_{s}$~\cite{Caron-Huot:2022eqs,Meltzer:2018tnm} as follows
\begin{equation}
    \begin{split}
        q^{2}_{0,s,matter}(z,\bz) ~=&~ g_s\,\frac{(-1)^s\,(s-1)!}{2}\,\int_{-\infty}^{+\infty}\,du\,\Bar{\Phi}(u,z,\bz)\,\pa_u^{s}\Phi(u,z,\bz) ~\propto~ {\cal E}_s(z,\bz) ~~,
    \end{split}
\end{equation}
where the higher-spin ANEC ${\cal E}_s$ takes the following form in momentum space
\begin{equation}
    {\cal E}_s(z,\bz) ~\propto~ \int_0^{\infty}\,d\o\,\o^s\,a^{\dagger}(\o,z,\bz)\,a(\o,z,\bz) ~~.
\end{equation}
For general $k$, the matter charge $q^{2}_{k,s,matter}$ becomes
\begin{equation}
   \begin{split}
      q^{2}_{k,s,matter}(z,\bz) ~=~ (-1)^{s}\,\frac{g_s}{2}\,\sum_{n=0}^{k}\,\frac{(-1)^{k-n}}{(k-n)!}&\,\frac{(s-1+n)!}{n!}\,\int_{-\infty}^{\infty}\, du\,u^{k-n}\\
      &\pa_z^{k-n}\,\Big[ \pa_u^{s}\,\Phi(u,z,\bz)\,\pa_z^n\,\pa_u^{-n}\Bar{\Phi}(u,z,\bz)\Big] ~~,
    \end{split}
\end{equation}
where the act of projecting onto individual modes amounts to smearing these operators on the full light sheet
\begin{equation}
    \begin{split}
        q^{k,s}_{(m,n),matter} ~=&~ \oint\,\frac{dz}{2\pi i}\,z^{\frac{s+k-1}{2}+m}\,\oint\,\frac{d\bz}{2\pi i}\,\bz^{-\frac{1-s+k}{2}+n}\,q^{2}_{k,s,matter}(z,\bz) ~~.
    \end{split}
    \label{equ:qks,mn,matter}
\end{equation}
With these operators represented completely in the matter sector, one interesting question is to ask if they form a closed algebra. For YM and gravity, we explicitly derived this in~\cite{Hu:2022txx} and section~\ref{sec:wedge}-\ref{sec:charges}.  One can perform the same calculation, evaluating the canonical bracket for generic (higher spin-$s$) operators. 
The first few terms obey
\begin{align}
    \Big[ q^2_{0,s}(z,\bz),\, q^2_{0,s'}(w,\bw)\Big] ~=&~ 0~~,\\
    \Big[ q^2_{0,s}(z,\bz),\, q^2_{1,s'}(w,\bw)\Big] ~=&~ i\,\frac{g_sg_{s'}}{g_{s+s'-2}}\,\frac{(s-1)!(s'-1)!}{(s+s'-3)!}\,\Big[ (s-1)\,\d^{(2)}(z-w)\,\pa_w\,q_{0,s+s'-2}(w,\bw)  \nonumber\\
    & \qquad\qquad\qquad\qquad ~+~(s+s'-1)\,\pa_w\d^{(2)}(z-w)\,q_{0,s+s'-2}(w,\bw)  \Big] ~~. \label{equ:HS-q0-q1}
\end{align}
Upon performing a mode expansion, (\ref{equ:HS-q0-q1}) yields
\begin{align}
    \Big[ q^{0,s}_{m,n},\, q^{1,s'}_{p,q}\Big] ~=&~ i\,\frac{g_sg_{s'}}{g_{s+s'-2}}\,\frac{(s-1)!(s'-1)!}{(s+s'-3)!}\,\Big[\, m\,s' - (s-1)\,p\,\Big]\,q^{0,s+s'-2}_{m+p,n+q}~~.
\end{align}
When $s=s'=2$ this reduces to the results in~\cite{Hu:2022txx} for gravity; however, one can see that for higher spin ($s,s'>2$) the assumption (\ref{equ:algebra-mode}) doesn't hold. Once we introduce a higher-$s$ operator, we have to include an infinite tower of them.  It will be very interesting to understand this from the perspective of collinear limits and the radial quantization brackets and also extend the analysis to incorporate color structure. We leave this for future investigation.

\paragraph{Matter Representations in 4D CFTs}
In connection with the conformal collider literature, it would be interesting to compute the pure matter realizations of the operator algebras we've constructed above explicitly in example 4D CFTs. In~\cite{Belin:2020lsr} it was observed that the expected symmetry algebra breaks down when computing commutators within correlation functions.  Here, this should be tied to the convergence of $u$, $z$, and $\bz$ integrals in these correlation functions. Along these lines, it would interesting to further understand the role of these light-sheet-supported operators in organizing the scattering observables. Indeed,~\cite{Caron-Huot:2022eqs} encountered a mixing between light-ray and light-sheet supported primaries in their investigations of detectors in weakly coupled theories.
Here we have seen that we need to couple to gauge fields to realize the symmetries in terms of light-ray operators that are local on the celestial sphere, which explains why these enhanced symmetries were not found in the cataloging of \cite{Cordova:2018ygx,Belin:2020lsr}.

\paragraph{Asymptotic Symmetries and the ${\cal S}$-matrix}
Finally, it is worthwhile to appreciate what we can take away from the phase space realization of the asymptotic symmetries when it comes to our understanding of the ${\cal S}$-matrix in gauge theory and gravity. We want to note that the task of identifying representations of the asymptotic symmetries on the matter phase space is an active area of exploration. See ex. \cite{Batlle:2017llu} for a discussion of the BMS algebra in 3D. Our discussions here point to how to extend this to the $w_{1+\infty}$ story. As compared to the discussion in the previous paragraph we are essentially restricting to the free 4D CFT. The underlying question is what symmetries of the free theory survive when we turn on interactions. While we can realize the celestial symmetries in terms of the  $in$ or $out$ phase space without talking about the equations of motion, we need the EOMs as soon as we want to tie these $in$ and $out$ contributions together and discuss symmetries of the full $\cal S$-matrix~\cite{Freidel:2021ytz,Freidel:2023gue}. In this language, the convergence of the operators we've constructed also effects whether or not the integer basis can sufficiently capture the radiative phase space in the interacting theory~\cite{Freidel:2022skz}.

\section*{Acknowledgements}
We would like to thank Luca Ciambelli, Joaquim Gomis, Mina Himwich, and especially Laurent Freidel for many useful conversations. The research of YH and SP is supported by the Celestial Holography Initiative at the Perimeter Institute for Theoretical Physics. This research program is supported by the Simons Collaboration on Celestial Holography. Research at the Perimeter Institute is supported by the Government of Canada through the Department of Innovation, Science and Industry Canada and by the Province of Ontario through the Ministry of Colleges and Universities.

\pagebreak

\appendix
\section{Useful Identities}\label{appen:IDs}
In this appendix, we collect some identities that are useful in our computations. 
\begin{itemize}
\item Delta function identity:
\begin{equation}
    \pa^n_x\,\d(x) ~=~ \frac{(-1)^n\,n!}{x^n}\,\d(x) ~~.
    \label{equ:delta-func-ID}
\end{equation}
\item Formal presentation of the Dirac delta function:
\begin{equation}
    \d(z-w) ~=~ z^{-1}\,\sum_k\,z^k\,w^{-k} ~~,~~\d(\bz-\bw) ~=~ \bz^{-1}\,\sum_k\,\bz^k\,\bw^{-k} ~~.
    \label{equ:deltafun-mode-exp}
\end{equation}
\item Derivative identity:
\begin{equation}
    \pa^p_{z_1}\,z_{12}^q ~=~ \frac{\Gamma(q+1)}{\Gamma(q-p+1)}\,z_{12}^{q-p} ~~.
    \label{equ:deriv-ID}
\end{equation}
\item Gamma function identity:
\begin{equation}
    \Gamma(\a-n) ~=~ (-1)^{n-1}\,\frac{\Gamma(-\a)\Gamma(1+\a)}{\Gamma(n+1-\a)} ~~,n\in\mathbb{Z}~.
    \label{equ:GammafuncID}
\end{equation}
\item Since the Mellin transform trades the energy $\o$ to the scaling dimension $\D$, we have the following identity
\begin{equation}
    (-\pa_{\o})^k\,\o^k ~:=~ (\D-1)_k ~~.
    \label{equ:omega-Delta-map}
\end{equation}
\item The Gauss hypergeometric function $_2F_1[a,b,c,1]$ can be expanded in terms of Gamma functions as follows 
\begin{equation}
    _2F_1[a,b,c,1] ~=~ \frac{\Gamma(c)\Gamma(c-a-b)}{\Gamma(c-a)\Gamma(c-b)}~~.
    \label{equ:2F1-exp}
\end{equation}
We will also use the regularized Gauss hypergeometric function defined as $_2\tilde{F}_1[a,b,c,z]=\, _2F_1[a,b,c,z]/\Gamma(c)$.
\item In our computations, we encounter the following summation several times 
\begin{equation}
    \begin{split}
        \sum_{n=0}^{k-l}\,(-k')_n\,\frac{(-1)^{n}}{n!} ~=&~ \sum_{n=0}^{k-l}\,\frac{(-1)^{n}}{n!}\frac{\Gamma(1-k')}{\Gamma(1-k'-n)} ~=~ \frac{(-1)^{k-l+1}\Gamma(1-k')}{k'\,(k-l)!\,\Gamma(l-k-k')}\\
        ~=&~ \frac{(-1)^{k-l+1}}{k'\,(k-l)!}\,\frac{\Gamma(k+k'-l+1)}{(-1)^{k-l-1}\,\Gamma(k')} ~=~ \frac{(k+k'-l)!}{k'!\,(k-l)!} ~~,
    \end{split}
    \label{equ:sum-ID}
\end{equation}
where we have used (\ref{equ:GammafuncID}). 
\end{itemize}

\section{Detector Operators for the Wedge Subalgebra}\label{appen:detector-op-wedge-examples}
In this appendix, we present several examples of the matter representation of wedge truncated celestial symmetry algebras. We will start with the case of gravity where $s=2$ in \ref{appen:detector-grav}. The $k=0$ and $k=1$ wedge modes belong to the global sector of the BMS, which will be shown explicitly. The first higher spin operator going beyond the BMS (i.e. $k=2$) will be presented as well, followed by similar discussions for Yang-Mills in \ref{appen:detector-YM}.

\subsection{Detector Operators in Gravity}\label{appen:detector-grav}
As we did in~\cite{Hu:2022txx}, let's consider complex scalar matter fields. The quadratic modes become 
\begin{equation}
    \begin{split}
        W^{k}_{(m,n),matter} ~=&~ \frac{1}{4}\,\sum_{l=0}^k(-1)^l\frac{(k-l+1)}{l!}\,\int_{-\infty}^{+\infty}du\,u^l\\
        &\qquad \oint\,\frac{dz\,d\bz}{(2\pi i)^2}\,z^{\frac{1+k}{2}+m}\,\bz^{\frac{1-k}{2}+n}\,\pa_z^l\Big[ \pa_u^2\Phi(u,z,\bz)\,\pa_z^{k-l}\pa_u^{-l}\Bar{\Phi}(u,z,\bz) \Big] ~~.
    \end{split}
\end{equation}
which form an infinite tower of higher spin charges corresponding to the wedge $w_{1+\infty}$
\begin{equation}
    \begin{split}
        \Big[W^{k}_{(m,n)},\,W^{k'}_{(p,q)} \Big] ~=~ \frac{i}{2}\,\Big[m(k'+1) - p(k+1)\Big]\,W^{k+k'-1}_{(m+p,n+q)} ~~\text{with}~~
        \begin{cases}
            -\frac{k+1}{2}\le m \le \frac{k+1}{2}~~,\\
            -\frac{k'+1}{2}\le p \le \frac{k'+1}{2} ~~.
        \end{cases}
    \end{split}
\end{equation}
Next, we are going to focus on the  special cases: $k=0$, $1$, and $2$.

\paragraph{$k=0$:} Wedge truncation requires $m=-\frac{1}{2}, \frac{1}{2}$ 
\begin{equation}
    \begin{split}
         W^{0}_{(-\frac{1}{2},n),matter} ~=&~ -\frac{1}{4}\,\oint\,\frac{dz}{2\pi i}\,\oint\,\frac{d\bz}{2\pi i}\,\bz^{\frac{1}{2}+n}\,{\cal E}_2(z,\bz) ~~, \\
        W^{0}_{(\frac{1}{2},n),matter} ~=&~ -\frac{1}{4}\,\oint\,\frac{dz}{2\pi i}\,z\,\oint\,\frac{d\bz}{2\pi i}\,\bz^{\frac{1}{2}+n}\,{\cal E}_2(z,\bz) ~~,\\ 
    \end{split}
\end{equation}
where ${\cal E}_2(z,\bz)$ is identified as the ANEC operator
\begin{equation}
   {\cal E}_2(z,\bz) ~=~  \int_{-\infty}^{+\infty}du\,\pa_u\Bar{\Phi}(u,z,\bz)\,\pa_u\Phi(u,z,\bz) ~~.
\end{equation}
Comparing with \cite{Cordova:2018ygx}, we see that $W^{0}_{(m,-\frac{1}{2}),matter}$ with $m=-\frac{1}{2},\frac{1}{2}$ correspond to the supertranslation charges ${\cal T}(f)$ for $f=1,z$ respectively. These two modes act exactly as the global translation generators
\begin{equation}
    \begin{split}
        \Big[ W^{0}_{(m,-\frac{1}{2})},\,{\cal O}_{\D,J}(z,\bz) \Big] ~=&~ \frac{i}{2}\,z^{m+\frac{1}{2}}\,{\cal O}_{\D+1,J}(z,\bz)~~,~~ m=-\frac{1}{2},\frac{1}{2}~~.
    \end{split}
    \label{equ:W0m-Pm-action}
\end{equation}

\paragraph{$k=1$:} The wedge truncation requires $m=-1, 0, 1$
\begin{equation}
    \begin{split}
         W^{1}_{(-1,n),matter} ~=&~ \frac{1}{8}\,\oint\,\frac{dz}{2\pi i}\,\oint\,\frac{d\bz}{2\pi i}\,\bz^{n}\, {\cal N}_{z}(z,\bz)  ~~, \\
        W^{1}_{(0,n),matter} ~=&~ \frac{1}{8}\,\oint\,\frac{dz}{2\pi i}\,\oint\,\frac{d\bz}{2\pi i}\,\bz^{n}\,\left[\frac{1}{2}\,{\cal L}_{-1}(z,\bz) + z\,{\cal N}_{z}(z,\bz) + \frac{1}{2}\,{\cal E}_{1}(z,\bz)\right] ~~,\\ 
        W^{1}_{(1,n),matter} ~=&~ \frac{1}{8}\,\oint\,\frac{dz}{2\pi i}\,z\,\oint\,\frac{d\bz}{2\pi i}\,\bz^{n}\,\Big[{\cal L}_{-1}(z,\bz) + z\,{\cal N}_{z}(z,\bz) + {\cal E}_{1}(z,\bz)\Big] ~~,\\ 
    \end{split}
\end{equation}
where recall that in \cite{Hu:2022txx} the spin-1 quadratic operator $W_{1,matter}(z,\bz)$ has been written in terms of a linear combination of the following generalized ANEC operators
\begin{equation}
    \begin{split}
        {\cal N}_{z}(z,\bz) ~=&~ \int_{-\infty}^{+\infty}du\,\pa_u\Phi(u,z,\bz)\,\pa_z\Bar{\Phi}(u,z,\bz) ~~,\\
        {\cal L}_{-1}(z,\bz) ~=&~ \int_{-\infty}^{+\infty}du\,u\,\pa_u\Phi(u,z,\bz)\,\pa_u\Bar{\Phi}(u,z,\bz) ~~,\\
        {\cal E}_{1}(z,\bz) ~=&~ \int_{-\infty}^{+\infty}du\,\Bar{\Phi}(u,z,\bz)\,\pa_u\Phi(u,z,\bz) ~~.
    \end{split}
\end{equation}
Comparing with \cite{Cordova:2018ygx}, we see that $W^{1}_{(m,0),matter}$ with $m=-1,0,1$ correspond to the superrotation charges ${\cal R}(Y^z)$ with $Y^z=1,z,z^2$, respectively. These three modes act exactly as the global ${\rm SL}(2)_L$ generators
\begin{equation}
    \begin{split}
        \Big[ W^{1}_{(m,0)},\,{\cal O}_{\D,J}(z,\bz) \Big] ~=&~ i\,z^{m}\,\Big[ (m+1)h + z\,\pa_z\Big]\,{\cal O}_{\D,J}(z,\bz)~~,~~ m=-1,0,1~~.
    \end{split}
    \label{equ:W1m-Lm-action}
\end{equation}

\paragraph{$k=2$:} The wedge truncation requires $m=-\frac{3}{2},-\frac{1}{2},\frac{1}{2},\frac{3}{2}$
\begin{equation}
    \begin{split}
        W^{2}_{(-\frac{3}{2},n),matter} ~=&~ \frac{3}{4}\,\oint\,\frac{dz}{2\pi i}\,\oint\,\frac{d\bz}{2\pi i}\,\bz^{-\frac{1}{2}+n}\, {\cal E}_{zz}(z,\bz) ~~,\\
        W^{2}_{(-\frac{1}{2},n),matter} ~=&~ \oint\,\frac{dz}{2\pi i}\,\oint\,\frac{d\bz}{2\pi i}\,\bz^{-\frac{1}{2}+n}\,\Bigg[ \frac{3}{4}\,z\,{\cal E}_{zz}(z,\bz)+ \frac{1}{2}\int_{-\infty}^{+\infty}du\,\Phi(u,z,\bz)\,\pa_z\Bar{\Phi}(u,z,\bz)\\
        & - \frac{1}{2}\int_{-\infty}^{+\infty}du\,u\,\pa_u\Phi(u,z,\bz)\,\pa_z\Bar{\Phi}(u,z,\bz) \Bigg] ~~,\\
       W^{2}_{(\frac{1}{2},n),matter} ~=&~ \oint\,\frac{dz}{2\pi i}\,\oint\,\frac{d\bz}{2\pi i}\,\bz^{-\frac{1}{2}+n}\,\Bigg[ \frac{3}{4}\,z^2\,{\cal E}_{zz}(z,\bz)+ z\,\int_{-\infty}^{+\infty}du\,\Phi(u,z,\bz)\,\pa_z\Bar{\Phi}(u,z,\bz)\\
        & - z\,\int_{-\infty}^{+\infty}du\,u\,\pa_u\Phi(u,z,\bz)\,\pa_z\Bar{\Phi}(u,z,\bz) \\
        & + \frac{1}{2}\,{\cal E}_0(z,\bz) + \frac{1}{2}\,\int_{-\infty}^{+\infty}du\,\Phi(u,z,\bz)\,\pa_u\Bar{\Phi}(u,z,\bz) - \frac{1}{4}\,{\cal L}_0(z,\bz)  \Bigg] ~~,\\ 
        W^{2}_{(\frac{3}{2},n),matter} ~=&~ \oint\,\frac{dz}{2\pi i}\,\oint\,\frac{d\bz}{2\pi i}\,\bz^{-\frac{1}{2}+n}\,\Bigg[ \frac{3}{4}\,z^3\,{\cal E}_{zz}(z,\bz)+ \frac{3}{2}\,z^2\,\int_{-\infty}^{+\infty}du\,\Phi(u,z,\bz)\,\pa_z\Bar{\Phi}(u,z,\bz)\\
        & - \frac{3}{2}\, z^2\,\int_{-\infty}^{+\infty}du\,u\,\pa_u\Phi(u,z,\bz)\,\pa_z\Bar{\Phi}(u,z,\bz) \\
        & + \frac{3}{2}\,z\,{\cal E}_0(z,\bz) + \frac{3}{2}\,z\,\int_{-\infty}^{+\infty}du\,\Phi(u,z,\bz)\,\pa_u\Bar{\Phi}(u,z,\bz) - \frac{3}{4}\,z\,{\cal L}_0(z,\bz)  \Bigg] ~~,\\ 
    \end{split}
\end{equation}
where 
\begin{equation}
    \begin{split}
        {\cal E}_{zz}(z,\bz) ~=&~ \int_{-\infty}^{+\infty}du\,\Phi(u,z,\bz)\,\pa^2_z\Bar{\Phi}(u,z,\bz) ~~,\\
        {\cal L}_{0}(z,\bz) ~=&~ \int_{-\infty}^{+\infty}du\,u^2\,\pa_u\Phi(u,z,\bz)\,\pa_u\Bar{\Phi}(u,z,\bz) ~~,\\
        {\cal E}_{0}(z,\bz) ~=&~ \int_{-\infty}^{+\infty}du\,\Phi(u,z,\bz)\,\Bar{\Phi}(u,z,\bz) ~~.
    \end{split}
\end{equation}
These operators do not have a standard BMS asymptotic symmetry interpretation. However, they still have the nice feature that they are constructed from smearing of local operators along the generators of null infinity, unlike the higher spin terms in the tower. 
The remaining elements of the wedge subalgebra can be generated from these.

\subsection{Detector Operators in Yang-Mills}\label{appen:detector-YM}

Again, we will consider complex scalar matter fields. We have the following quadratic modes
\begin{equation}
    \begin{split}
        J^{a,k}_{(m,n),matter} ~=&~ -\frac{i}{2}\,[T^a]_{\b\g}\,\sum_{l=0}^k\,\frac{(-1)^l}{l!}\,\int_{-\infty}^{+\infty}du\,u^l\\
        &\qquad \oint\,\frac{dz\,d\bz}{(2\pi i)^2}\,z^{\frac{k}{2}+m}\,\bz^{-\frac{k}{2}+n}\,\pa_z^l\Big[ \pa_u\Phi^{\b}(u,z,\bz)\,\pa_z^{k-l}\pa_u^{-l}\Bar{\Phi}^{\g}(u,z,\bz) \Big] ~~,
    \end{split}
\end{equation}
which form an infinite tower of higher spin charges corresponding to the wedge S algebra
\begin{equation}
    \begin{split}
        \Big[J^{a,k}_{(m,n)},\,J^{b,k'}_{(p,q)} \Big] ~=~ if^{abc}\,J^{c,k+k'}_{(m+p,n+q)} ~~\text{with}~~
        \begin{cases}
            -\frac{k}{2}\le m \le \frac{k}{2}~~,\\
            -\frac{k'}{2}\le p \le \frac{k'}{2} ~~.
        \end{cases}
    \end{split}
\end{equation}

\paragraph{$k=0$:} The wedge truncation requires $m=0$ 
\begin{equation}
    J^{a,0}_{(0,n),matter} ~=~ \oint\,\frac{dz\,d\bz}{(2\pi i)^2}\,\bz^{n}\, J^{2,a}_{0,matter}(z,\bz)  ~~,
\end{equation}
where $J^{2,a}_{0,matter}(z,\bz)$ is presented in (\ref{equ:charge-flux-op-matter}). Compare with \cite{Cordova:2018ygx}, the mode $J^{a,0}_{(0,0),matter}$ corresponds to the charge operator ${\cal Q}^a(g)$ with $g=1$
\begin{equation}
    \Big[J^{a,0}_{(0,0)},\, {\cal O}^b_{\D,J}(z,\bz)\Big] ~=~ if^{abc}\,{\cal O}^c_{\D,J}(z,\bz) ~~.
\end{equation}

\paragraph{$k=1$:} The wedge truncation requires $m=-\frac{1}{2},\frac{1}{2}$
\begin{equation}
    \begin{split}
        J^{a,1}_{(-\frac{1}{2},n),matter} ~=&~ -\frac{i}{2}\,[T^a]_{\b\g}\,\oint\,\frac{dz\,d\bz}{(2\pi i)^2}\,\bz^{-\frac{1}{2}+n}\,\int_{-\infty}^{+\infty}du\,\pa_u\Phi^{\b}(u,z,\bz)\,\pa_z\Bar{\Phi}^{\g}(u,z,\bz) ~~,\\
        J^{a,1}_{(\frac{1}{2},n),matter} ~=&~ -\frac{i}{2}\,[T^a]_{\b\g}\,\oint\,\frac{dz\,d\bz}{(2\pi i)^2}\,z\,\bz^{-\frac{1}{2}+n}\,\int_{-\infty}^{+\infty}du\,\pa_u\Phi^{\b}(u,z,\bz)\,\pa_z\Bar{\Phi}^{\g}(u,z,\bz)\\
        &~-\frac{i}{2}\,[T^a]_{\b\g}\,\oint\,\frac{dz\,d\bz}{(2\pi i)^2}\,\bz^{-\frac{1}{2}+n}\,\int_{-\infty}^{+\infty}du\,u\,\pa_u\Phi^{\b}(u,z,\bz)\,\pa_u^{-1}\Bar{\Phi}^{\g}(u,z,\bz)~~.
    \end{split}
\end{equation}

\section{Additional Computational Details}\label{appen:derivations}

In this appendix, we flesh out some derivations that were omitted in the main text to streamline our story. 

\subsection{Quadratic Operators}\label{appen:[q2s,O]}
Here we explicitly show that the quadratic operator
\begin{equation}
   \begin{split}
      q^{2,\a}_{k,s_1(s_2)}(z,\bz) ~=&~ \frac{g^{\a\b\g}_{12p}}{2}\frac{1}{(2\pi)^3}\,\sum_{n=0}^{k}\,\frac{(-1)^{k-n}}{(k-n)!}\frac{(s_1-1+n)!}{n!}\,\int_0^{\infty}d\o_1\,\int_0^{\infty}d\o_2\,(-i\o_2)^{s_1-n}\\
        &\qquad \left(\frac{\o_1}{\o_2}\right)^{s_1-s_2}\,(-i\pa_{\o_1})^{k-n}\d(\o_1-\o_2)\,\pa_z^{k-n}\,\Big[ a^{\dagger,\b}_{s_2}(\o_1,z,\bz)\,\pa_z^n\,a^{\g}_{s_2}(\o_2,z,\bz)\Big] 
    \end{split}
\end{equation}
satisfies (\ref{equ:[q2s,a]}). 

\paragraph{$\Big[q^{2,\a}_{k,s_1(s_2)}(z_1,\bz_1),\,a^{\b}_{s_2}(\o,z_2,\bz_2)\Big]$} Let's start with the commutator between the quadratic operator and the annihilation operator. A straightforward computation yields
\begin{equation}
    \begin{split}
        \Big[q^{2,\a}_{k,s_1(s_2)}(z_1,\bz_1),\,a^{\b}_{s_2}(\o,z_2,\bz_2)\Big] 
         ~=&~ -g^{\a\b\g}_{12p}\,\sum_{n=0}^{k}\,\frac{(-1)^{k-n}}{(k-n)!}\frac{(s_1-1+n)!}{n!}\,\o^{s_1-s_2-1}\,(-i\pa_{\o})^{k-n}\\
        &\qquad (-i\o)^{s_1-n}\left(\frac{1}{\o}\right)^{s_1-s_2}\,
        \pa_{z_1}^{k-n}\,\d^{(2)}(z_{12})\,\pa^n_{z_2}\,a^{\g}_{s_2}(\o,z_2,\bz_2) ~~.\\
    \end{split}
    \label{equ:[q2k,a]-omega}
\end{equation}
The $\o$-terms can be manipulated as follows 
\begin{equation}
    \begin{split}
        & \o^{s_1-s_2-1}\,(-i\pa_{\o})^{k-n}\,(-i\o)^{s_1-n}\left(\frac{1}{\o}\right)^{s_1-s_2}
        ~=~ (-i)^{s_1-k}\,\o^{s_1-s_2-1}\,(-\pa_{\o})^{k-n}\o^{k-n}\o^{s_2-k}\\
        ~=&~ (-i)^{s_1-k}\,\o^{s_1-s_2-1}\,(\D-1)_{k-n}\,\o^{s_2-k}
        ~=~ (-i)\,(\D+s_1-s_2-2)_{k-n}\,(-i\o)^{s_1-k-1} ~~,
    \end{split}
    \label{equ:omega-manipulation}
\end{equation}
where we have used the identity (\ref{equ:omega-Delta-map}).
Then the commutator becomes
\begin{equation}
    \begin{split}
         \Big[q^{2,\a}_{k,s_1(s_2)}(z_1,\bz_1),\,a^{\b}_{s_2}(\o,z_2,\bz_2)\Big]
         ~=&~ ig^{\a\b\g}_{12p}\,\sum_{n=0}^{k}\,\frac{(-1)^{k-n}}{(k-n)!}\frac{(s_1-1+n)!}{n!}\,(\D+s_1-s_2-2)_{k-n} \\
         &\qquad\qquad~~ (-i\o)^{s_1-k-1}\,\pa_{z_1}^{k-n}\,\d^{(2)}(z_{12})\,\pa^n_{z_2}\,a^{\g}_{s_2}(\o,z_2,\bz_2) ~~,\\
    \end{split}
\end{equation}
which is exactly the second equation in (\ref{equ:[q2s,a]}).

\paragraph{$\Big[q^{2,\a}_{k,s_1(s_2)}(z_1,\bz_1),\,a^{\dagger,\b}_{s_2}(\o,z_2,\bz_2)\Big]$} This computation is somewhat more involved compared to the previous one since, to write everything in terms of oscillators located at $(z_2,\bz_2)$, we have to re-distribute the derivatives as follows
\begin{equation}
    \begin{split}
       & \pa_{z_1}^{k-n}\,\Bigg( a^{\dagger,\g}_{s_2}(\o_1,z_1,\bz_1)\,\pa_{z_1}^n\,\Big[ a^{\d}_{s_2}(\o_2,z_1,\bz_1),\,a^{\dagger,\b}_{s_2}(\o,z_2,\bz_2)\Big]\Bigg) \\
       ~=&~ 2(2\pi)^3\,\d^{\b\d}\,\frac{\d(\o-\o_2)}{\o}\,\sum_{r=0}^n\begin{pmatrix}
       n\\
       r
       \end{pmatrix}(-1)^r\,\pa_{z_2}^r a^{\dagger,\g}_{s_2}(\o_1,z_2,\bz_2)\,\pa_{z_1}^{k-r}\d^{(2)}(z_{12})~~.
    \end{split}
\end{equation}
Then, direct computation yields
\begin{equation}
    \begin{split}
       & \Big[q^{2,\a}_{k,s_1(s_2)}(z_1,\bz_1),\,a^{\dagger,\b}_{s_2}(\o,z_2,\bz_2)\Big] 
       ~=~ g^{\a\g\b}_{12p}\,\sum_{n=0}^{k}\sum_{r=0}^n\,\frac{(-1)^{k-n}}{(k-n)!}\frac{(s_1-1+n)!}{n!}\,\o^{s_2-s_1-1}\,(-i\o)^{s_1-n}\\
        &\qquad\qquad\qquad\qquad\qquad\qquad(i\pa_{\o})^{k-n} \o^{s_1-s_2}\,\begin{pmatrix}
       n\\
       r
       \end{pmatrix}(-1)^r\,
        \pa_{z_1}^{k-r}\,\d^{(2)}(z_{12})\,\pa^r_{z_2}\,a^{\dagger,\g}_{s_2}(\o,z_2,\bz_2) ~~.\\
    \end{split}
\end{equation}
Note that one can exchange the order of the sums $\sum_{n=0}^{k}\sum_{r=0}^n = \sum_{r=0}^{k}\sum_{n=r}^k $ and similar to what was done in (\ref{equ:omega-manipulation}),  the $\o$-terms become
\begin{equation}
    \o^{s_2-s_1-1}\,(-i\o)^{s_1-n}\,(i\pa_{\o})^{k-n} \o^{s_1-s_2} ~=~ i\,(-1)^{n+s_1}\,(\D+s_2-n-2)_{k-n}\,(i\o)^{s_1-k-1} ~~.
\end{equation}
Then the commutator becomes
\begin{equation}
    \begin{split}
        & \Big[q^{2,\a}_{k,s_1(s_2)}(z_1,\bz_1),\,a^{\dagger,\b}_{s_2}(\o,z_2,\bz_2)\Big] 
       ~=~ (-1)^{s_1}\,i\,g^{\a\g\b}_{12p}\,\sum_{r=0}^{k}\frac{(-1)^{k-r}}{r!}\,\sum_{n=r}^k\,\frac{(s_1-1+n)!}{(k-n)!(n-r)!}\\
       &\qquad\qquad\qquad\qquad  (\D+s_2-n-2)_{k-n}\,(i\o)^{s_1-k-1}\,\pa_{z_1}^{k-r}\,\d^{(2)}(z_{12})\,\pa^r_{z_2}\,a^{\dagger,\g}_{s_2}(\o,z_2,\bz_2) ~~.\\
    \end{split}
\end{equation}
The $n$-sum can be done as follows. 
\begin{equation}
    \begin{split}
   & \sum_{n=r}^k\,\frac{(s_1-1+n)!(\D+s_2-n-2)_{k-n}}{(k-n)!(n-r)!} \\
    ~=&~ \frac{\Gamma(r+s_1)\Gamma(\D-r+s_2-1)}{\Gamma(1+k-r)\Gamma(\D-k+s_2-1)}\,_2F_1[r-k,r+s_1,2-\D+r-s_2,1] \\
    ~=&~ \frac{(s_1+r-1)!}{(k-r)!}\,(\D+s_1+s_2-2)_{k-r}~~,
    \end{split}
\end{equation}
where we have used the Gamma function expansion of the Gauss hypergeometric function (\ref{equ:2F1-exp})
and the Gamma function identity (\ref{equ:GammafuncID}) twice. Explicitly, 
\begin{equation}
    \begin{split}
       & \Gamma(\D-r+s_2-1) ~=~ (-1)^{r-k}\frac{\Gamma(2-\D+k-s_2)\Gamma(\D-k+s_2-1)}{\Gamma(2-\D+r-s_2)} ~~,\\
       & \Gamma(\D+s_1+s_2-1) ~=~ (-1)^{r-k}\frac{\Gamma(2-\D-s_1-s_2-r+k)\Gamma(\D+s_1+s_2+r-k-1)}{\Gamma(2-\D-s_1-s_2)}~~.
    \end{split}
\end{equation}
Finally plugging this sum result back into the commutator, we obtain the first equation in (\ref{equ:[q2s,a]}).

\subsection{Recursive Proof of the S algebra}\label{appen:proof-YM}
In this section, we show that the action (\ref{equ:quad-op-bracket}) guarantees that if the S algebra holds for a given pair of $(k,k')$, it will hold for $(k+1,k')$ and $(k,k'+1)$ with the following assumptions: 1. the action (\ref{equ:quad-op-bracket}) satisfies the Jacobi identity (\ref{equ:JacobiID}); 2. the algebra holds when $(k,k')=(1,k')$ and $(k,k')=(k,1)$. This proof will be the S algebra analog of appendix C in \cite{Himwich:2021dau} and holds without the wedge truncation assumption. We will discuss the wedge truncation in appendix~\ref{appen:Jacobi-wedge}. 

First, the Jacobi identity implies
\begin{equation}
    \Big[ J^{a,k}_{m,n},\,\Big[J^{b,k'}_{p,q},\,{\cal O}^{\g}_{\D,J}\Big]\Big] - \Big[ J^{b,k'}_{p,q},\,\Big[J^{a,k}_{m,n},\,{\cal O}^{\g}_{\D,J}\Big]\Big] 
        ~=~ i\,f^{abc}\,\Big[J^{c,k+k'}_{m+p,n+q},\,{\cal O}^{\g}_{\D,J}\Big]~.
\end{equation}
We then consider acting with another $J^{d,1}_{j,l}$ on both sides of this equation
\begin{equation}\resizebox{0.9\textwidth}{!}{$%
        \Big[ J^{d,1}_{j,l},\,\Big[ J^{a,k}_{m,n},\,\Big[J^{b,k'}_{p,q},\,{\cal O}^{\g}_{\D,J}\Big]\Big]\Big] - \Big[ J^{d,1}_{j,l},\,\Big[ J^{b,k'}_{p,q},\,\Big[J^{a,k}_{m,n},\,{\cal O}^{\g}_{\D,J}\Big]\Big]\Big] 
        ~=~ i\,f^{abc}\,\Big[ J^{d,1}_{j,l},\,\Big[J^{c,k+k'}_{m+p,n+q},\,{\cal O}^{\g}_{\D,J}\Big]\Big]~.$}%
    \label{equ:J1cov-YM}
\end{equation}
Applying manipulations similar to those in \cite{Himwich:2021dau}, we have
\begin{equation}
    \begin{split}
        \Big[ J^{d,1}_{j,l},\,\Big[ J^{a,k}_{m,n},\,\Big[J^{b,k'}_{p,q},\,{\cal O}^{\g}_{\D,J}\Big]\Big]\Big] ~=&~ {\cal D}^{1,d\g\d}_{j,l}(h)\,\Big[ J^{a,k}_{m,n},\,\Big[J^{b,k'}_{p,q},\,{\cal O}^{\d}_{\D-1,J}\Big]\Big] \\
        &~+~ if^{dbe}\,\Big[ J^{a,k}_{m,n},\,\Big[J^{e,k'+1}_{p+j,q+l},\,{\cal O}^{\g}_{\D,J}\Big]\Big]\\
        &~+~ if^{dae}\,\Big[ J^{e,k+1}_{m+j,n+l},\,\Big[J^{b,k'}_{p,q},\,{\cal O}^{\g}_{\D,J}\Big]\Big]\\
        \Big[ J^{d,1}_{j,l},\,\Big[ J^{b,k'}_{p,q},\,\Big[J^{a,k}_{m,n},\,{\cal O}^{\g}_{\D,J}\Big]\Big]\Big] ~=&~  {\cal D}^{1,d\g\d}_{j,l}(h)\,\Big[ J^{b,k'}_{p,q},\,\Big[J^{a,k}_{m,n},\,{\cal O}^{\d}_{\D-1,J}\Big]\Big] \\
        &~+~ if^{dae}\,\Big[ J^{b,k'}_{p,q},\,\Big[J^{e,k+1}_{m+j,n+l},\,{\cal O}^{\g}_{\D,J}\Big]\Big]\\
        &~+~ if^{dbe}\,\Big[ J^{e,k'+1}_{p+j,q+l},\,\Big[J^{a,k}_{m,n},\,{\cal O}^{\g}_{\D,J}\Big]\Big]\\
        \Big[ J^{d,1}_{j,l},\,\Big[J^{c,k+k'}_{m+p,n+q},\,{\cal O}^{\g}_{\D,J}\Big]\Big] ~=&~ {\cal D}^{1,d\g\d}_{j,l}(h)\,\Big[ J^{c,k+k'}_{m+p,n+q},\,{\cal O}^{\d}_{\D-1,J}\Big] \\
        &~+~ if^{dce}\,\Big[ J^{e,k+k'+1}_{m+p+j,n+q+l},\,{\cal O}^{\g}_{\D,J}\Big]\Big]
    \end{split}
\end{equation}
where we have used the assumption that the algebra holds when $(k,k')=(1,k')$ and $(k,k')=(k,1)$.
Note that because
\begin{equation}
    f^{abc}f^{dce} ~=~ f^{dbc}f^{ace} ~+~ f^{dac}f^{cbe}
\end{equation}
(\ref{equ:J1cov-YM}) becomes
\begin{equation}\resizebox{0.9\textwidth}{!}{$%
    \begin{aligned}
      & 0~=~ {\cal D}^{1,d\g\d}_{j,l}(h)\,\Bigg\{\, \Big[ J^{a,k}_{m,n},\,\Big[J^{b,k'}_{p,q},\,{\cal O}^{\g}_{\D,J}\Big]\Big] - \Big[ J^{b,k'}_{p,q},\,\Big[J^{a,k}_{m,n},\,{\cal O}^{\g}_{\D,J}\Big]\Big] 
        - i\,f^{abc}\,\Big[J^{c,k+k'}_{m+p,n+q},\,{\cal O}^{\g}_{\D,J}\Big]\,\Bigg\} \\
    &+if^{dbc}\,\Bigg\{\,\Big[ J^{a,k}_{m,n},\,\Big[J^{c,k'+1}_{p+j,q+l},\,{\cal O}^{\g}_{\D,J}\Big]\Big] - \Big[ J^{c,k'+1}_{p+j,q+l},\,\Big[J^{a,k}_{m,n},\,{\cal O}^{\g}_{\D,J}\Big]\Big] - if^{ace}\,\Big[ J^{e,k+k'+1}_{m+p+j,n+q+l},\,{\cal O}^{\g}_{\D,J}\Big]\Big]\,\Bigg\} \\
    &+if^{dac}\,\Bigg\{\,\Big[ J^{c,k+1}_{m+j,n+l},\,\Big[J^{b,k'}_{p,q},\,{\cal O}^{\g}_{\D,J}\Big]\Big] - \Big[ J^{b,k'}_{p,q},\,\Big[J^{c,k+1}_{m+j,n+l},\,{\cal O}^{\g}_{\D,J}\Big]\Big] - if^{cbe}\,\Big[ J^{e,k+k'+1}_{m+p+j,n+q+l},\,{\cal O}^{\g}_{\D,J}\Big]\Big]\,\Bigg\} \\
    \end{aligned} $}%
\end{equation}
where the first line is our starting point and it is zero, therefore the second and the third line have to be zero as well. Namely, the algebra holds for $(k+1,k')$ and $(k,k'+1)$. 

Importantly, this analysis doesn't rely on anything about the wedge truncation. One can also show that for Yang-Mills, we only need to explicitly check the algebra for $(k,k')=(0,0)$, $(0,1)$, $(1,1)$, and we can get the full tower based on this recursive argument.

\subsubsection{Wedge Truncation}\label{appen:Jacobi-wedge}

Let's write down the differential operator (\ref{equ:Differential-op}) for the S algebra generator explicitly
\begin{equation}
    \begin{split}
        {\cal D}_{m,n}^{k,a\b\g}(h_2) ~=&~ i^2\,[T^a]_{\b\g}\,\sum_{l=0}^k\begin{pmatrix}
      m+\frac{k}{2}\\
      l
      \end{pmatrix}(2h_2-1)_l\, z^{\frac{k}{2}+m-l}\,\bz^{-\frac{k}{2}+n}\,\pa_z^{k-l}
    \end{split}
    \label{equ:diff-YM-matter}
\end{equation}
where to be general, we consider the charged operator to transform in an unspecified representation $[T^a]_{\b\g}$ of the gauge group.
The Jacobi identity (\ref{equ:JacobiID}) reduces to 
\begin{equation}
    {\cal D}_{p,q}^{k',b\g\d}(h_2)\,{\cal D}_{m,n}^{k,a\d\e}\left(h_2-\frac{k'}{2}\right) - {\cal D}_{m,n}^{k,a\g\d}(h_2)\,{\cal D}_{p,q}^{k',b\d\e}\left(h_2-\frac{k}{2}\right) ~=~ i\,f^{abd}\,{\cal D}_{m+p,n+q}^{k+k',d\g\e}(h_2) ~~.
    \label{equ:cond-diff-YM-matter}
\end{equation}
Plugging in (\ref{equ:diff-YM-matter}), and after some algebra, we have
\begin{equation}
    [T^bT^a]_{\g\e}\,{\cal A}(k',p;k,m) ~-~ [T^aT^b]_{\g\e}\,{\cal A}(k,m;k',p) ~=~ i^3\,f^{abd}\,[T^d]_{\g\e}\,{\cal B}(k+k';m+p)~~,
    \label{equ:YM-Jacobi-simplified}
\end{equation}
where
\begin{equation}
    \begin{split}
       &{\cal A}(k',p;k,m) ~=~ \sum_{l=0}^{k'}\sum_{j=0}^k\sum_{r=0}^{k'-l}\begin{pmatrix}
            p+\frac{k'}{2}\\
            l\\
        \end{pmatrix}\begin{pmatrix}
            m+\frac{k}{2}\\
            j
        \end{pmatrix}\begin{pmatrix}
            k'-l\\
            r
        \end{pmatrix}\,(2h_2-1)_l\,(2h_2-k'-1)_j\\
        &\qquad\qquad\qquad\qquad
        \left(\frac{k}{2}+m-j\right)_r\,z^{\frac{k+k'}{2}+m+p-l-j-r}\,\pa_z^{k+k'-l-r-j} ~~,\\
        &{\cal B}(k+k';m+p) ~=~ \sum_{r=0}^{k+k'}\begin{pmatrix}
            m+p+\frac{k+k'}{2}\\
            r
        \end{pmatrix}(2h_2-1)_r\,z^{\frac{k+k'}{2}+m+p-r}\,\pa_z^{k+k'-r}~~.
    \end{split}
\end{equation}
Note that 
\begin{equation}
    [T^bT^a]_{\g\e} ~-~ [T^aT^b]_{\g\e} ~=~ i^3\,f^{abd}\,[T^d]_{\g\e}~~.
\end{equation}
Below we will show that in the wedge truncation, namely $m\in[-\frac{k}{2},\frac{k}{2}]$ and $p\in[-\frac{k'}{2},\frac{k'}{2}]$, we have
\begin{equation}
    {\cal A}(k',p;k,m) ~=~ {\cal A}(k,m;k',p) ~=~ {\cal B}(k+k';m+p)~~.
\end{equation}
Since we have already shown that we only need to check the $k=0$ and $k=1$ cases explicitly for all other cases hold recursively, we will examine these two special cases below.

\paragraph{Warmup: $k=k'=0$ Subalgebra}
First, look at the special case $k=k'=0$. Then (\ref{equ:YM-Jacobi-simplified}) becomes
\begin{equation}
    {[T^bT^a]_{\g\e}}\,\begin{pmatrix}
            p\\
            0\\
        \end{pmatrix}\begin{pmatrix}
            m\\
            0
        \end{pmatrix}\,z^{m+p} ~-~ {[T^aT^b]_{\g\e}}\,\begin{pmatrix}
            m\\
            0\\
        \end{pmatrix}\begin{pmatrix}
            p\\
            0
        \end{pmatrix}\,z^{m+p} ~=~ {i^3\,f^{abd}\,[T^d]_{\g\e}}\,\begin{pmatrix}
            m+p\\
            0
        \end{pmatrix}\,z^{m+p}
\end{equation}
which holds for arbitrary $m$ and $p$.

\paragraph{Special case: $k=1$, $m=-\frac{1}{2}$ and all $k'$ with $p\in[-\frac{k'}{2},\frac{k'}{2}]$}
In this case, we have $m+\frac{k}{2}=0$, so that ${\cal A}(k',p;k,m)$ and ${\cal B}(k+k';m+p)$ are equal, while ${\cal A}(k,m;k',p)$ takes a more complicated form
\begin{equation}
    \begin{split}
        {\cal A}(k',p;k,m) ~=&~ \sum_{l=0}^{k'}\begin{pmatrix}
            p+\frac{k'}{2}\\
            l
        \end{pmatrix}\,(2h_2-1)_l\,z^{\frac{k'}{2}+p-l}\,\pa_z^{k'+1-l} ~=~ {\cal B}(k+k';m+p)~~,\\
        {\cal A}(k,m;k',p) ~=&~ \sum_{l=0}^{k'}\sum_{r=0}^1\,\begin{pmatrix}
            p+\frac{k'}{2}\\
            l
        \end{pmatrix}\,(2h_2-2)_l\,\left(\frac{k'}{2}+p-l\right)_r\,z^{\frac{k'}{2}+p-l-r}\,\pa_z^{k'+1-l-r} ~~.\\
    \end{split}
\end{equation}
For ${\cal A}(k,m;k',p)$, we can reorganize and simplify the expression as follows
\begin{equation}
    \begin{split}
       & {\cal A}(k,m;k',p) 
        ~=~ \sum_{l=0}^{k'}\begin{pmatrix}
            p+\frac{k'}{2}\\
            l
        \end{pmatrix}\frac{\Gamma(2h_2-1)}{\Gamma(2h_2-1-l)}z^{\frac{k'}{2}+p-l}\,\pa_z^{k'+1-l}\\
        &\qquad\qquad\qquad\qquad ~+~ \sum_{l=0}^{k'}\frac{(p+\frac{k'}{2})!}{l!(p+\frac{k'}{2}-l-1)!}\frac{\Gamma(2h_2-1)}{\Gamma(2h_2-1-l)}\,z^{\frac{k'}{2}+p-l-1}\,\pa_z^{k'-l}\\
        ~=&~ \sum_{l=0}^{k'}\begin{pmatrix}
            p+\frac{k'}{2}\\
            l
        \end{pmatrix}\frac{\Gamma(2h_2-1)}{\Gamma(2h_2-1-l)}z^{\frac{k'}{2}+p-l}\,\pa_z^{k'+1-l}
        ~+~ \sum_{l=0}^{k'+1}\,l\,\begin{pmatrix}
            p+\frac{k'}{2}\\
            l
        \end{pmatrix}\frac{\Gamma(2h_2-1)}{\Gamma(2h_2-l)}\,z^{\frac{k'}{2}+p-l}\,\pa_z^{k'-l+1}\\
        ~=&~ \sum_{l=0}^{k'}\begin{pmatrix}
            p+\frac{k'}{2}\\
            l
        \end{pmatrix}\,\frac{\Gamma(2h_2-1)}{\Gamma(2h_2-l)}\,\Big[ (2h_2-l-1) + l \Big]\,z^{\frac{k'}{2}+p-l}\,\pa_z^{k'+1-l} ~=~ {\cal A}(k',p;k,m)~~.
    \end{split}
\end{equation}

\paragraph{Using SL(2) covariance:} We will now show that if the  Jacobi identity holds for fixed $k$ and $m$, then it holds for all $m\in[-\frac{k}{2},\frac{k}{2}]$. Then, together with the result of the special case for $k=1$ and $m=-\frac{1}{2}$ that we just computed above, we finish the proof of the wedge S-algebra for $k=1$. 
Recall that the SL(2) generator $L_m$ with $m=0,\pm 1$ acts on a primary as follows
\begin{equation}
    \Big[ L_m,\,{\cal O}_{h,\bh}(z,\bz)\Big] ~=~ z^m\,\Big[ (m+1)\,h + z\pa_z\Big]\,{\cal O}_{h,\bh}(z,\bz) ~:=~ {\cal L}_m(h)\,{\cal O}_{h,\bh}(z,\bz)~~.
    \label{equ:Lm-O}
\end{equation}
Doing mode expansion on both sides yields
\begin{equation}
    \Big[ L_m,\,{\cal O}_{h,\bh,(p,q)}(\bz)\Big] ~=~ \Big[ m\,(h-1)-n\Big]\,{\cal O}_{h,\bh,(m+p,q)}(\bz)~~.
\end{equation}
Given that the charge $J^a_k$ is a primary with $h=\frac{2+k}{2}$, we have
\begin{equation}
    \Big[ L_j,\,J^{a,k}_{m,n}\Big] ~=~ \frac{1}{2}\Big[ j\,k-2\,m\Big]\,J^{a,k}_{m+j,n} ~~.
    \label{equ:Lj-J-action}
\end{equation}
Start with the assumption that Jacobi holds for a fixed $k$, a fixed $m\in [-\frac{k}{2},\frac{k}{2}]$, and all $k'$ and $p\in [-\frac{k'}{2},\frac{k'}{2}]$
\begin{equation}
    \begin{split}
        \Big[ J^{a,k}_{m,n},\,\Big[J^{b,k'}_{p,q},\,{\cal O}^{\g}_{\D,J}\Big]\Big] - \Big[ J^{b,k'}_{p,q},\,\Big[J^{a,k}_{m,n},\,{\cal O}^{\g}_{\D,J}\Big]\Big] 
        ~=&~ i\,f^{abc}\,\Big[J^{c,k+k'}_{m+p,n+q},\,{\cal O}^{\g}_{\D,J}\Big]~~,
    \end{split}
\end{equation}
then act $L_j$ on both sides, 
\begin{equation}
    \begin{split}
        \Big[ L_j,\,\Big[ J^{a,k}_{m,n},\,\Big[J^{b,k'}_{p,q},\,{\cal O}^{\g}_{\D,J}\Big]\Big]\Big] - \Big[ L_j,\,\Big[ J^{b,k'}_{p,q},\,\Big[J^{a,k}_{m,n},\,{\cal O}^{\g}_{\D,J}\Big]\Big]\Big] ~=~ i\,f^{abc}\,\Big[ L_j,\,\Big[J^{c,k+k'}_{m+p,n+q},\,{\cal O}^{\g}_{\D,J}\Big]\Big]~~.
    \end{split}
    \label{equ:SL2cov-YM}
\end{equation}
Using (\ref{equ:quad-op-bracket}), (\ref{equ:Lm-O}), and (\ref{equ:Lj-J-action}) repeatedly,  (\ref{equ:SL2cov-YM}) becomes
\begin{equation}\resizebox{0.9\textwidth}{!}{$%
    \begin{aligned}
        &0~=~{\cal L}_j(h)\,\Bigg\{\,\Big[ J^{a,k}_{m,n},\,\Big[J^{b,k'}_{p,q},\,{\cal O}^{\g}_{\D,J}\Big]\Big] + \Big[ J^{b,k'}_{p,q},\,\Big[J^{a,k}_{m,n},\,{\cal O}^{\g}_{\D,J}\Big]\Big] - i\,f^{abc}\,\Big[ J^{c,k+k'}_{m+p,n+q},\,{\cal O}^{\g}_{\D,J}\Big] \,\Bigg\}\\
        &+\frac{1}{2}\,\Big(jk-2m\Big)\,\Bigg\{\,\Big[ J^{a,k}_{m+j,n},\,\Big[J^{b,k'}_{p,q},\,{\cal O}^{\g}_{\D,J}\Big]\Big] - \Big[ J^{b,k'}_{p,q},\,\Big[J^{a,k}_{m+j,n},\,{\cal O}^{\g}_{\D,J}\Big]\Big]- i\,f^{abc}\, \Big[J^{c,k+k'}_{m+p+j,n+q},\,{\cal O}^{\g}_{\D,J}\Big] \,\Bigg\}\\
        &+\frac{1}{2}\,\Big(jk'-2p\Big)\,\Bigg\{\, \Big[ J^{a,k}_{m,n},\,\Big[J^{b,k'}_{p+j,q},\,{\cal O}^{\g}_{\D,J}\Big]\Big] - \Big[ J^{b,k'}_{p+j,q},\,\Big[J^{a,k}_{m,n},\,{\cal O}^{\g}_{\D,J}\Big]\Big] - i\,f^{abc}\, \Big[J^{c,k+k'}_{m+p+j,n+q},\,{\cal O}^{\g}_{\D,J}\Big]\,\Bigg\}~~.\\
    \end{aligned}$}%
\end{equation}
The first and the third lines are zero based on our assumptions. Therefore, 
\begin{equation}
    \begin{split}\Big(jk-2m\Big)\,\Bigg\{\,\Big[ J^{a,k}_{m+j,n},\,\Big[J^{b,k'}_{p,q},\,{\cal O}^{\g}_{\D,J}\Big]\Big] &~-~ \Big[ J^{b,k'}_{p,q},\,\Big[J^{a,k}_{m+j,n},\,{\cal O}^{\g}_{\D,J}\Big]\Big] \\
    &\qquad ~-~ i\,f^{abc}\, \Big[J^{c,k+k'}_{m+p+j,n+q},\,{\cal O}^{\g}_{\D,J}\Big] \,\Bigg\} ~=~ 0
    \end{split}
\end{equation}
which tells us
\begin{enumerate}
    \item when $m<\frac{k}{2}$, taking $j=+1$, and $2m-k\ne 0$, the Jacobi identity holds for $m+1$;\\
    \item when $m>-\frac{k}{2}$, taking $j=-1$, and $2m+k\ne 0$, the Jacobi identity holds for $m-1$;\\
    \item when $m=\pm\frac{k}{2}$, namely $m$ is exactly on the boundary of the wedge truncation, we cannot show Jacobi holds for $m\pm 1$. In other words, we cannot go beyond the wedge truncation using SL(2) covariance arguments.
\end{enumerate}

\subsubsection{Example: Failure of the Jacobi beyond the Wedge}\label{appen:YMexample-failJacobi}

In order to have a better understanding of whether or not the commutator of the quadratic operators closes beyond the wedge truncation, let us look at a simple example: $k=0$, $k'=1$ in Yang-Mills. In this case, (\ref{equ:YM-Jacobi-simplified}) becomes
\begin{equation} \resizebox{0.9\textwidth}{!}{$%
    \begin{aligned}
       &[T^bT^a]_{\g\e}\, \sum_{l=0}^1\begin{pmatrix}
        p+\frac{1}{2}\\
        l
        \end{pmatrix}(2h_2-1)_l\,z^{\frac{1}{2}+p-l}\pa_z^{1-l}\,z^m - [T^aT^b]_{\g\e}\,\sum_{l=0}^1\begin{pmatrix}
        p+\frac{1}{2}\\
        l
        \end{pmatrix}(2h_2-1)_l\,z^{\frac{1}{2}+p-l+m}\pa_z^{1-l}\\
         &~=~ i^3\,f^{abd}\,[T^d]_{\g\e}\, \sum_{l=0}^1\begin{pmatrix}
        p+\frac{1}{2}+m\\
        l
        \end{pmatrix}(2h_2-1)_l\,z^{\frac{1}{2}+p-l+m}\pa_z^{1-l}  ~~.
    \end{aligned} $}%
\end{equation}
We see that the above equation holds only when $m=0$.

\subsection{Linear Truncation of the Algebra}\label{appen:linear-truncation}

\paragraph{Derivation of (\ref{equ:[q2k,q1k']})}\label{appen:[q2k,q1k']}
We split this computation into two steps: 1. computing the commutator between $q^{2,\a}_{k,s}$ and $H^{\b}_{k',s}$; and 2. acting with $\pa^{k'+s}_{z_2}$ on both sides.
Note that both the $q^{2,\a}_{k,s}$ and $H^{\b}_{k',s}$ operators split into two parts: one only containing $a_s$ and $a^{\dagger}_s$ and the other only containing $b_s$ and $b^{\dagger}_s$. Moreover, these two parts share a similar structure and are mutually commuting. In what follows we will omit the $b_s$ and $b^{\dagger}_s$ terms since the commutators for this part can be computed in a similar manner. 
Using (\ref{equ:delta-func-ID}) and (\ref{equ:[q2k,a]-omega}), direct computation yields
\begin{equation}
    \begin{split}
       & \Big[q^{2,\a}_{k,s}(z_1,\bz_1),H^{\b}_{k',s}(z_2,\bz_2)\Big] ~=~ -\frac{1}{2}\,\frac{i^{k'}}{(2\pi)}\,\int_0^{\infty} d\o\,\o^{1-k'}\,\d(\o)\,  \Big[q^{2,\a}_{k,s}(z_1,\bz_1),a_s^{\b}(\o,z_2,\bz_2) \Big] \\
        ~=&~  \frac{g^{\a\b\g}_{12p}}{2}\frac{i^{k'+k-s}}{(2\pi)}\,\sum_{l=0}^{k}\,\frac{(-1)^{k-l}}{(k-l)!}\frac{(s_1-1+l)!}{l!}\,\int_0^{\infty} d\o\,\o^{-k'}\,\d(\o)\,\int_0^{\infty}d\o_2\,\o_2^{s-l}\\
        &\qquad\qquad\qquad\qquad\qquad\qquad\qquad\qquad \pa_{\o_2}^{k-l}\d(\o-\o_2)\, \pa_{z_1}^{k-l}\,\d^{(2)}(z_{12})\,\pa^l_{z_2}\,a^{\g}_{s}(\o_2,z_2,\bz_2) ~~. \\
    \end{split}
\end{equation}
For the $\o$- and $\o_2$-terms we implement the following manipulations
\begin{equation}
    \begin{split}
        \o^{-k'}\,\o_2^{s-l}\,\pa_{\o_2}^{k-l}\d(\o-\o_2) ~=&~ \sum_{n=0}^{k-l}\,\begin{pmatrix}
        k-l\\
        n
        \end{pmatrix}\,\Big(\pa_{\o_2}^n\o_2^{-k'}\Big)\,\Big(\pa_{\o_2}^{k-l-n}\d(\o-\o_2)\Big)\,\o_2^{s-l}\\
        ~=&~ (-1)^{k-l}\,(k-l)!\,\o_2^{-k-k'+s}\,\d(\o-\o_2)\,\sum_{n=0}^{k-l}\,(-k')_n\,\frac{(-1)^{n}}{n!}~~.
    \end{split}
\end{equation}
Moreover, notice that the $n$-sum has been already done in (\ref{equ:sum-ID}).
Altogether,
\begin{equation}
    \begin{split}
        &\Big[q^{2,\a}_{k,s}(z_1,\bz_1),H^{\b}_{k',s}(z_2,\bz_2)\Big] \\
        ~=&~ i\,g^{\a\b\g}_{12p}\,\sum_{l=0}^{k}\,\frac{(s_1-1+l)!}{l!}\,\begin{pmatrix}
        k+k'-l\\
        k'
        \end{pmatrix}\pa_{z_1}^{k-l}\,\d^{(2)}(z_{12})\,\pa^l_{z_2}\,H^{\g}_{k+k'+1-s,s}(z_2,\bz_2)~~.
    \end{split}
    \label{equ:[q2s,H]}
\end{equation}
Next, acting with $\pa^{k'+s}_{z_2}$ on both sides of equation (\ref{equ:[q2s,H]}) yields
\begin{equation}
    \begin{split}
      & \Big[q^{2,\a}_{k,s}(z_1,\bz_1),\,q^{1,\b}_{k',s}(z_2,\bz_2)\Big] ~=~ i\,g^{\a\b\g}_{12p}\,\sum_{n=0}^{k}\,\frac{(s-1+k-n)!}{(k-n)!}\,\begin{pmatrix}
        k'+n\\
        k'
        \end{pmatrix}\,\sum_{m=0}^{s+k'}\,\begin{pmatrix}
        s+k'\\
        m
        \end{pmatrix}\\
        &\qquad\qquad\qquad\qquad\qquad\qquad\qquad\qquad (-1)^m\, \pa_{z_1}^{n+m}\d^{(2)}(z_{12})\,\pa_{z_2}^{s-1-m-n}\,q^{1,\g}_{k+k'+1-s,s}(z_2,\bz_2) \\
        &~=~ i\,g^{\a\b\g}_{12p}\,\sum_{p=0}^{k+k'+s}\,\sum_{n=max[0,p-s-k']}^{min[k,p]}\,\frac{(s-1+k-n)!}{(k-n)!}\,\begin{pmatrix}
        k'+n\\
        k'
        \end{pmatrix}\,\begin{pmatrix}
        s+k'\\
        p-n
        \end{pmatrix}\,(-1)^{p-n}\\ &\qquad\qquad\qquad\qquad\qquad\qquad\qquad\qquad\qquad\qquad\qquad \pa_{z_1}^{p}\d^{(2)}(z_{12})\,\pa_{z_2}^{s-1-p}\,q^{1,\g}_{k+k'+1-s,s}(z_2,\bz_2)~~,
    \end{split}
\end{equation}
which reduces to (\ref{equ:[q2k,q1k']}) after defining the coefficient $C^{(s)}(k,k';p)$ as in (\ref{equ:C(s)(k,k')-def}).

\paragraph{\texorpdfstring{Simplification of $\Big[J^{1,a}_{k}(z_1,\bz_1),\,J^{2,b}_{k'}(z_2,\bz_2)\Big]$}{Simplification of [J{1,a}_{k}(z_1,bz_1),\,J{2,b}_{k'}(z_2,bz_2)]}}\label{appen:tildeC}

Given (\ref{equ:[q1k,q2k']}) and \eqref{equ:C(1)} we have
\begin{equation}
    \begin{split}
        \Big[J^{1,a}_{k}(z_1,\bz_1),\,J^{2,b}_{k'}(z_2,\bz_2)\Big] ~=&~ if^{abc}\,\delta^{(2)}(z_{12})\,J^{1,c}_{k+k'}(z_2,\bz_2) \\
        &~+~ if^{abc}\,\sum_{n=0}^{k+k'+1}\,\tilde{C}(k',k;n)\,\pa_{z_1}^{n}\delta^{(2)}(z_{12})\,\pa^{-n}_{z_2}\,J^{1,c}_{k+k'}(z_2,\bz_2) ~~, \\
    \end{split}
    \label{equ:[J1a,J2b]-simplify}
\end{equation}
where we've defined
\begin{equation}
    \begin{split}
       \tilde{C}(k',k;n) ~:=&~ (-1)^n\,\sum_{p=max[k'+1,n]}^{k+k'+1}\,\begin{pmatrix}
        p\\
        n
        \end{pmatrix}(-1)^{p+k'}\,\begin{pmatrix}
        p-1\\
        k'
        \end{pmatrix}\,\begin{pmatrix}
        k+k'+1\\
        p
        \end{pmatrix} ~~.
    \end{split}
    \label{equ:tildeC}
\end{equation}
The lower bound of the $p$-sum is a max-function. To further simplify this expression we need to discuss the range of $n$. First, let's consider $n\le k'$. The $p$-sum becomes $\sum_{p=k'+1}^{k+k'+1}$, and we have
\begin{equation}
    \begin{split}
        \tilde{C}(k',k;n\le k') ~=&~ (-1)^{n+1}\,\frac{\Gamma(2+k+k')}{\Gamma(1+n)\Gamma(1+k)}\,_2\tilde{F}_1[-k,1+k',2-n+k',1] \\
        ~=&~ (-1)^{n+1}\,\frac{\Gamma(2+k+k')}{\Gamma(1+n)\Gamma(1+k)}\,\frac{\Gamma(1+k-n)}{\Gamma(2+k+k'-n)\Gamma(1-n)} ~~,
    \end{split}
    \label{equ:tildeC-1}
\end{equation}
where $_2\tilde{F}_1$ is the regularized Gauss hypergeometric function. 
When $n\ge k'+1$, the $p$-sum becomes $\sum_{p=n}^{k+k'+1}$, and we similarly have 
\begin{equation}
    \begin{split}
        \tilde{C}(k',k;n\ge k'+1) ~=&~ \frac{(-1)^{k'}\,\Gamma(2+k+k')}{n\Gamma(1+k')\Gamma(2+k+k'-n)}\,_2\tilde{F}_1[n,-1+n-k-k',n-k',1] \\
        ~=&~ (-1)^{k'}\,\frac{\Gamma(2+k+k')}{n\Gamma(1+k')\Gamma(2+k+k'-n)}\,\frac{\Gamma(1+k-n)}{\Gamma(-k')\Gamma(k+1)} ~~.
    \end{split}
    \label{equ:tildeC-2}
\end{equation}
To further simplify these expressions, we can analyze the poles coming from Gamma functions, and then use the identity (\ref{equ:GammafuncID}). This simplification can be divided into the following three cases. 
\begin{enumerate}
    \item When $n=0$, (\ref{equ:tildeC-1}) simply reduces to
\begin{equation}
    \tilde{C}(s',s;0) ~=~ -1 ~~.
\end{equation}
    \item When $1\le n\le k$, in both (\ref{equ:tildeC-1}) and (\ref{equ:tildeC-2}) there is an infinity in the denominators while the numerators are finite. Therefore, 
    \begin{equation}
        \tilde{C}(k',k;1\le n\le k) ~=~ 0 ~~.
    \end{equation}
    \item When $k+1 \le n\le k+k'+1$, in (\ref{equ:tildeC-1}) and (\ref{equ:tildeC-2}) both the numerators and denominators diverge. We can apply (\ref{equ:GammafuncID}) to simplify them, finding
    \begin{equation}
        \begin{split}
          & k+1 \le  n\le k':~~\tilde{C}(k',k;n) ~=~  (-1)^{n}\,\frac{\Gamma(2+k+k')}{n\,\Gamma(1+n)\Gamma(1+k)}\,\frac{\Gamma(1+k-n)}{\Gamma(2+k+k'-n)\Gamma(-n)} \\
          &\qquad\qquad\qquad\qquad\qquad\qquad ~=~  (-1)^{n+k+1}\,\frac{(1+k+k')!}{n\,k!\,(k+k'+1-n)!(n-k-1)!}~~, \\
          &  k'+1,k+1 \le  n\le k+k'+1:~~\tilde{C}(k',k;n) ~=~ \frac{(-1)^{k-n-1}\,(k+k'+1)!}{n\,(k+k'+1-n)!k!(n-k-1)!}~~, 
        \end{split}
    \end{equation}
    where we have used the following equations
    \begin{equation}
        \begin{split}
        \Gamma(1+n)\,\Gamma(-n) ~=&~ (-1)^{k-1}\,\Gamma(n-k)\,\Gamma(k+1-n) ~~,\\
            \Gamma(1+k')\,\Gamma(-k') ~=&~ (-1)^{k+k'-n-1}\,\Gamma(n-k)\,\Gamma(k+1-n) ~~.
        \end{split}
    \end{equation}
    Note that in these two regions, $\tilde{C}(k',k;n)$ gives the same result, therefore
    \begin{equation}
        \begin{split}
            k+1 \le  n\le k+k'+1:~~\tilde{C}(k',k;n) ~=&~ (-1)^{k+n+1}\, \begin{pmatrix}
        k+k'+1\\
        n
        \end{pmatrix}\,\begin{pmatrix}
        n-1\\
        k
        \end{pmatrix} ~~.
        \end{split}
    \end{equation}
\end{enumerate}
Finally, $\Big[J^{1,a}_{k}(z_1,\bz_1),\,J^{2,b}_{k'}(z_2,\bz_2)\Big]$ can be simplified as follows
\begin{equation}
    \begin{split}
       & \Big[J^{1,a}_{k}(z_1,\bz_1),\,J^{2,b}_{k'}(z_2,\bz_2)\Big] ~=~ if^{abc}\,\delta^{(2)}(z_{12})\,J^{1,c}_{k+k'}(z_2,\bz_2) ~-~if^{abc}\,\delta^{(2)}(z_{12})\,J^{1,c}_{k+k'}(z_2,\bz_2)  \\
        &\qquad\qquad\qquad\qquad\qquad\qquad ~+~ if^{abc}\,\sum_{n=k+1}^{k+k'+1}\,\tilde{C}(k',k;n)\,\pa_{z_1}^{n}\delta^{(2)}(z_{12})\,\pa^{-n}_{z_2}\,J^{1,c}_{k+k'}(z_2,\bz_2)\\
        &~=~if^{abc}\,\sum_{n=k+1}^{k+k'+1}\,(-1)^{k+n+1}\, \begin{pmatrix}
        k+k'+1\\
        n
        \end{pmatrix}\,\begin{pmatrix}
        n-1\\
        k
        \end{pmatrix}\,\pa_{z_1}^{n}\delta^{(2)}(z_{12})\,\pa^{-n}_{z_2}\,J^{1,c}_{k+k'}(z_2,\bz_2) ~~. \\
    \end{split}
\end{equation}

\subsection{Operators from the Opposite Helicity Sector}\label{appen:check-tildeq}
In this appendix, we explicitly show that the solution we presented in (\ref{equ:general-tilde-q}) indeed satisfies the conditions in (\ref{equ:condition-tildeq}).
Unlike the other operators we have encountered in this paper, for $\widetilde{q}^{2,\a}_{k,s}(z,\bz)$, the expression in terms of oscillators doesn't split:
\begin{equation}
    \begin{split}
        \widetilde{q}^{2,\a}_{k,s}(z,\bz) 
    ~=&~ \frac{1}{4}\frac{g^{\a\g\d}_{12p}}{(2\pi)^3}\,\sum_{l=0}^{k}\,\frac{(-1)^{k-l}}{(k-l)!}\frac{(s-1+l)!}{l!}\,\int_0^{\infty}d\o_1\,\int_0^{\infty}d\o_2\\
    &\Bigg\{\, (-i\o_2)^{-s-l}\,(-i\pa_{\o_1})^{k-l}\d(\o_1-\o_2)\, \pa_{z}^{k-l}\Big[\, a^{\dagger,\g}_s(\o_1,z,\bz)\,\pa_z^l b^{\d}_s(\o_2,z,\bz)\,\Big]\\ 
     &~+~ (i\o_2)^{-s-l}\,(i\pa_{\o_1})^{k-l}\d(\o_1-\o_2)\, \pa_{z}^{k-l}\Big[\, b^{\g}_s(\o_1,z,\bz)\,\pa_z^l a^{\dagger,\d}_s(\o_2,z,\bz)\,\Big]\,\Bigg\} ~~.\\ 
    \end{split}
\end{equation}
The commutators between $\Tilde{q}^{2,a}_s(z,\bz)$ and the oscillators are as follows
\begin{equation}
    \begin{split}
        &\Big[ \Tilde{q}^{2,\a}_{k,s}(z,\bz),\, b_s^{\dagger,\b}(\o,w,\bw)\Big] \\
        ~=&~ \frac{g^{\a\g\b}_{12p}}{2}\,\sum_{l=0}^{k}\,\frac{(-1)^{k-l}}{(k-l)!}\frac{(s-1+l)!}{l!}\,\int_0^{\infty}d\o_1\,\frac{1}{\o}(-i\o)^{-s-l}\,(-i\pa_{\o_1})^{k-l}\d(\o_1-\o)\\
    &\qquad\qquad \sum_{n=0}^l
    \begin{pmatrix}
    l\\
    n
    \end{pmatrix}(-1)^n\,\pa_w^n a^{\dagger,\g}_s(\o_1,w,\bw)\,\pa_z^{k-n}\d^{(2)}(z-w)\\
     &~+~ \frac{g^{\a\b\g}_{12p}}{2}\,\sum_{l=0}^{k}\,\frac{(-1)^{k-l}}{(k-l)!}\frac{(s-1+l)!}{l!}\,\int_0^{\infty}d\o_2\,\frac{1}{\o}(i\o_2)^{-s-l}\,(i\pa_{\o})^{k-l}\d(\o-\o_2)\\
    &\qquad\qquad \pa_w^l a^{\dagger,\g}_s(\o_2,w,\bw)\,\pa_z^{k-l}\d^{(2)}(z-w) ~~,\\
    \end{split}
    \label{equ:[tildeq2k,bdagger]}
\end{equation}
and 
\begin{equation}
    \begin{split}
        & \Big[ \Tilde{q}^{2,\a}_{k,s}(z,\bz),\, a_s^{\b}(\o,w,\bw)\Big] \\
        ~=&~ -\frac{g^{\a\b\g}_{12p}}{2}\,\sum_{l=0}^{k}\,\frac{(-1)^{k-l}}{(k-l)!}\frac{(s-1+l)!}{l!}\,\int_0^{\infty}d\o_2\,\frac{1}{\o}(-i\o_2)^{-s-l}\,(-i\pa_{\o})^{k-l}\d(\o-\o_2)\\
    &\qquad\qquad \pa_w^l b^{\g}_s(\o_2,w,\bw)\,\pa_z^{k-l}\d^{(2)}(z-w)\\
     &~-~ \frac{g^{\a\g\b}_{12p}}{2}\,\sum_{l=0}^{k}\,\frac{(-1)^{k-l}}{(k-l)!}\frac{(s-1+l)!}{l!}\,\int_0^{\infty}d\o_1\,\frac{1}{\o}(i\o)^{-s-l}\,(i\pa_{\o_1})^{k-l}\d(\o_1-\o)\\
    &\qquad\qquad \sum_{n=0}^l
    \begin{pmatrix}
    l\\
    n
    \end{pmatrix}(-1)^n\,\pa_w^n b^{\g}_s(\o_1,w,\bw)\,\pa_z^{k-n}\d^{(2)}(z-w) ~~.\\
    \end{split}
    \label{equ:[tildeq2k,a]}
\end{equation}
Note that the commutator can be split as follows
\begin{equation}
    \Big[ \widetilde{q}^{2,\a}_{k,s}(z,\bz), H^{\b}_{k',s}(w,\bw) \Big] ~=~ \Big[ \widetilde{q}^{2,\a}_{k,s}(z,\bz), H^{\b,+}_{k',s}(w,\bw) \Big]  ~+~ \Big[ \widetilde{q}^{2,\a}_{k,s}(z,\bz), H^{\b,-}_{k',s}(w,\bw) \Big] ~~,
\end{equation}
where $H^{\b,+}_{k',s}(w,\bw)$ only contains $b^{\dagger,\b}_s$ and $H^{\b,-}_{k',s}(w,\bw)$ only contains $a^{\b}_s$. 
For the commutator with $H^{\b,+}_{k',s}(w,\bw)$, using (\ref{equ:[tildeq2k,bdagger]}) we have
\begin{equation}
    \begin{split}
        &\Big[ \widetilde{q}^{2,\a}_{k,s}(z,\bz), H^{\b,+}_{k',s}(w,\bw) \Big] ~=~ \frac{g_{12p}^{\a\g\b}}{4}\frac{(-1)^{3k'+2k+1-s}i^{k+k'-s}}{(2\pi)}\,\sum_{n=0}^{k}\,\sum_{r=0}^n\frac{(s-1+n)!}{(k-n)!n!}\begin{pmatrix}
        n\\
        r
        \end{pmatrix}(-1)^r\\
        &\qquad\qquad \int_0^{\infty}d\o_1\,\int_0^{\infty}d\o\,\o^{-s-n-k'}\,\d(\o)\,\pa_{\o_1}^{k-n}\d(\o_1-\o)\,\pa_w^r\,a^{\dagger,\g}_{s}(\o_1,w,\bw)\,\pa_z^{k-r}\d^{(2)}(z-w) \\
       &~+~  \frac{g_{12p}^{\a\b\g}}{4}\frac{(-1)^{3k'+k+1}i^{k+k'-s}}{(2\pi)}\,\sum_{n=0}^{k}\,\frac{(s-1+n)!}{(k-n)!n!}\\
       &\qquad\qquad \int_0^{\infty}d\o\,\o^{-k'}\,\d(\o)\,\int_0^{\infty}d\o_2\,\o_2^{-s-n}\,\pa_{\o}^{k-n}\d(\o-\o_2)\,\pa_w^n\,a^{\dagger,\g}_{s}(\o_2,w,\bw)\,\pa_z^{k-n}\d^{(2)}(z-w) \\
    \end{split}
\end{equation}
where the first term can be simplified as follows
\begin{equation}
   \begin{split}
        \text{1st term} 
        ~=&~ \frac{i\,g^{\a\g\b}_{12p}}{2i^{2s}}\,\sum_{r=0}^k\,\frac{(r+s-1)!}{r!}\,\begin{pmatrix}
    k+k'-r\\
    k'
    \end{pmatrix} \pa_w^r\,\widetilde{H}^{\g,-}_{k+k'+1-s,s}(w,\bw)\,\pa_z^{k-r}\d^{(2)}(z-w)~~.
   \end{split}
\end{equation}
For the second term, we can integrate out the $\o$ integral as follows
\begin{equation}
\begin{split}
   & \int_0^{\infty}d\o\,\d(\o)\,\o^{-k'}\,\pa_{\o}^{k-n}\d(\o-\o_2)\\
   ~=&~ \int_0^{\infty}d\o\,\d(\o)\,(-1)^{k-n}\,\sum_{m=0}^{k-n}\begin{pmatrix}
        k-n\\
        m
        \end{pmatrix}\,(-k')_m\,\o_2^{-k'-m}\,\pa_{\o_2}^{k-n-m}\d(\o-\o_2)\\
        ~=&~ \sum_{m=0}^{k-n}(-1)^m\,\begin{pmatrix}
        k-n\\
        m
        \end{pmatrix}\,(-k')_m\,(k-n-m)!\o_2^{-k'-k+n}\,\d(\o_2)\\
        ~=&~ (k-n)!\,\o_2^{-k'-k+n}\,\d(\o_2)\,\sum_{m=0}^{k-n}(-1)^m\,\frac{(-k')_m}{m!} ~=~ (k-n)!\,\o_2^{-k'-k+n}\,\d(\o_2)\,\begin{pmatrix}
        k+k'-n\\
        k'
        \end{pmatrix} ~~,
\end{split}
\label{equ:omega-integeral-q2sRs'}
\end{equation}
where in the last line we used the identity (\ref{equ:sum-ID}). 
Therefore
\begin{equation}
    \begin{split}
       & \text{2nd term} ~=~ \frac{g_{12p}^{\a\b\g}}{4}\frac{(-1)^{3k'+k+1}i^{k+k'-s}}{(2\pi)}\,\sum_{n=0}^{k}\,\frac{(s-1+n)!}{n!}\,\begin{pmatrix}
        k+k'-n\\
        k'
        \end{pmatrix}\\ &\qquad\qquad\qquad\qquad\qquad~~~\int_0^{\infty}d\o_2\,\o_2^{-s-k-k'}\,\d(\o_2)\, \pa_w^l a^{\dagger,\g}_s(\o_2,w,\bw)\,\pa_z^{k-n}\d^{(2)}(z-w) \\
       & ~=~ \frac{i\,g_{12p}^{\a\b\g}}{2(-1)^{2k}i^{4s}}\,\sum_{n=0}^{k}\,\frac{(s-1+n)!}{n!}\,\begin{pmatrix}
        k+k'-n\\
        k'
        \end{pmatrix}\,\pa_w^n\,\widetilde{H}^{\g,-}_{k+k'+1-s,s}(w,\bw)\,\pa_z^{k-n}\d^{(2)}(z-w)~~.
    \end{split}
\end{equation}
Altogether, we have
\begin{equation}
    \begin{split}
         \Big[ \widetilde{q}^{2,\a}_{k,s}(z,\bz), H^{\b,+}_{k',s}(w,\bw) \Big] ~=&~ \frac{i}{2}\,\Big( i^{2s}\,g^{\a\g\b}_{12p} + g_{12p}^{\a\b\g}\Big)\,\sum_{n=0}^{k}\,\frac{(s-1+n)!}{n!}\,\begin{pmatrix}
        k+k'-n\\
        k'
        \end{pmatrix}\\
       &\qquad\qquad\qquad \pa_w^n\,\widetilde{H}^{\g,-}_{k+k'+1-s,s}(w,\bw)\,\pa_z^{k-n}\d^{(2)}(z-w)~~.
    \end{split}
    \label{equ:[tildeq2,H+]}
\end{equation}
Similarly, the commutator with $H^{\b,-}_{k',s}(w,\bw)$ takes the following form
\begin{equation}
    \begin{split}
         \Big[ \widetilde{q}^{2,\a}_{k,s}(z,\bz), H^{\b,-}_{k',s}(w,\bw) \Big] ~=&~ \frac{i}{2}\,\Big(  g_{12p}^{\a\b\g} + i^{2s}\,g^{\a\g\b}_{12p}\Big)\,\sum_{n=0}^{k}\,\frac{(s-1+n)!}{n!}\,\begin{pmatrix}
        k+k'-n\\
        k'
        \end{pmatrix}\\
       &\qquad\qquad\qquad \pa_w^n\,\widetilde{H}^{\g,+}_{k+k'+1-s,s}(w,\bw)\,\pa_z^{k-n}\d^{(2)}(z-w)~~.
    \end{split}
    \label{equ:[tildeq2,H-]}
\end{equation}
Finally, summing (\ref{equ:[tildeq2,H+]}) and (\ref{equ:[tildeq2,H-]}), we obtain
\begin{equation}
    \begin{split}
        \Big[ \widetilde{q}^{2,\a}_{k,s}(z,\bz), H^{\b}_{k',s}(w,\bw) \Big] 
       ~=&~ \frac{i}{2}\,\Big(  g_{12p}^{\a\b\g} + i^{2s}\,g^{\a\g\b}_{12p}\Big)\,\sum_{n=0}^{k}\,\frac{(s-1+n)!}{n!}\,\begin{pmatrix}
        k+k'-n\\
        k'
        \end{pmatrix}\\
      &\qquad\qquad\qquad \pa_w^n\,\widetilde{H}^{\g}_{k+k'+1-s,s}(w,\bw)\,\pa_z^{k-n}\d^{(2)}(z-w) ~~.\\
    \end{split}
\end{equation}
Note that when $s=1$, we have $g_{12p}^{\a\b\g}\sim f^{abc}$, and the prefactor becomes
\begin{equation}
    \frac{i}{2}\,\Big(  f^{abc} + i^{2}\,f^{acb} \Big) ~=~ i\,f^{abc} ~~,
\end{equation}
while when $s=2$, there is no internal symmetry, and the prefactor reduces to 
\begin{equation}
    \frac{i}{2}\,\Big(  g_{12p} + i^{4}\,g_{12p}\Big) ~=~ i\,g_{12p} ~~.
\end{equation}

\subsubsection{Example Cubic Operator: $\widetilde{J}^{3,a}_1$ in Yang-Mills}\label{appen:example-Jtilde}

With the quadratic operators (\ref{equ:J2-gluon}) and (\ref{equ:tildeJ}), we can evaluate the quadratic truncation of the bracket (\ref{equ:q-tildeq-alg}) with $k=0$ and $k'=1$. First, direct computation yields
\begin{equation}
    \begin{split}
        \Big[ J^{2,a}_{0}(z,\bz),&\,\widetilde{J}^{2,b}_1(w,\bw)\Big] ~=~ i\,f^{abc}\,d^{(2)}(z-w)\, \widetilde{J}^{2,a}_1(w,\bw)\\
        & ~-~ \frac{i}{2(2\pi)^3}\,\pa_w\d^{(2)}(z-w)\,\Big\{ f^{ace}f^{bdc}\,\int_0^{+\infty}d\o_1\,\frac{1}{\o_1^2}\,a^{\dagger,d}_1(\o_1,w,\bw)\,b^e_1(\o_1,w,\bw)\\
        &~+~ f^{abc}f^{cde}\,\int_0^{+\infty}d\o_1\,\int_0^{+\infty}d\o_2\,\frac{1}{\o_2}\,\pa_{\o_1}\d(\o_1-\o_2)\,a^{\dagger,d}_1(\o_1,w,\bw)\,b^e_1(\o_2,w,\bw)\,\Big\}~~.
    \end{split}
\end{equation}
The anomalous term should be canceled by the bracket $\Big[ J^{1,a}_{0}(z,\bz),\,\widetilde{J}^{3,b}_1(w,\bw)\Big]$ which leads to 
\begin{equation}
    \begin{split}
        \widetilde{J}^{3,a}_1(w,\bw) ~=&~ \frac{i^2}{2(2\pi)^6}\,f^{bcd}f^{def}\,\int_{-\infty}^{+\infty}du\,\int_0^{+\infty}d\o_1\,a^{\dagger,c}_1(\o_1,w,\bw)\,e^{i\o_1u}\\
        &\int_{+\infty}^udu'\,\int_0^{+\infty}d\o_2\,a^{\dagger,e}_1(\o_2,w,\bw)\,e^{i\o_2u'}\,\int_0^{+\infty}d\o_3\,\frac{1}{\o_3}\,b_1^f(\o_3,w,\bw)e^{-i\o_3u'}~~.
    \end{split}
\end{equation}
Again, as expected, $\widetilde{J}^{3,a}_1$ only contains $a^{\dagger,a}_1$ and $b^a_1$, therefore the third line of (\ref{equ:q-tildeq-alg}) always holds.

\bibliographystyle{utphys}
\bibliography{references}

\end{document}